\documentstyle[amstex,12pt,epsf]{article}
\newcommand{\fnote}[2]{\begingroup
\def\thefootnote{#1}\footnote{#2}
\addtocounter{footnote}{-1}\endgroup}
\newcommand{\email}{mabe@@th.phys.titech.ac.jp}

\newcommand{\nn}{\nonumber}
\newcommand{\ph}{\phantom{-}}
\newcommand{\po}{\phantom{1}}
\newcommand{\I}{\text{I}}

\newcommand{\III}{\text{III}}
\newcommand{\IV}{\text{IV}}
\newcommand{\V}{\text{V}}
\begin{document}
\pagestyle{empty}
\begin{flushright}{\tt hep-th/0109155}\\
TIT/HEP-471\\
September, 2001
\end{flushright}
\vspace{18pt}

\begin{center}
{\Large  Dualities between K3 fibered Calabi--Yau
 three-folds   } \\ 
\vspace{4pt}
\vspace{16pt}

Mitsuko Abe\fnote{*}{\email}       
 
\vspace{16pt}
{\sl Faculty of Engineering\\
Shibaura  Institute of Technology \\ 
Minato-ku, Tokyo 108-8548, Japan\\}

\center {and}

{\sl Department of Physics\\
Tokyo Institute of Technology \\ 
Oh-okayama, Meguro, Tokyo 152-8551, Japan\\}

\end{center}
\vspace{1cm}
\begin{abstract}
We propose a way to examine  N=1 and N=2 string   
dualities  on Calabi-Yau three-folds and their  extensions.  
Our way is to find out or to construct two
types of toric representations  of 
a  Calabi-Yau three-fold, which contain    
phases  topologically equivalent or phases connected by flops. 
We  discuss  how to find relations among  Calabi-Yau 
three-folds  realized in different toric representations. 
We examine several examples  of Calabi-Yau three-folds 
that have  the  Hodge numbers, 
$(h^{1,1},h^{2,1})\!=\!(5,185)$ and  the various numbers of K3 fibers. 
We observe that  each phase of  our examples  
 contains Del Pezzo 4-cycles,   $B_8$ in six  ways.     

\end{abstract}

\vfill
\pagebreak

\pagestyle{plain}
\setcounter{page}{1}
\setcounter{footnote}{2}

\baselineskip=16pt
\section{Introduction}
\par 
A motivation of our work  
is to  examine N=1 and N=2 string  dualities   
 from   identification of   Calabi-Yau three-folds (CY3s). 
We propose to utilize   
 several different types of   toric representations, 
i.e., local coordinates of a  CY3,
which are topologically equivalent.
\par
There are two  points that  characterize  a toric 
representation in the above case:  
 one is  the existence of extra     
 tensor multiplets in  6-dimensional  intermediate stage and the other  
the existence of double   K3 fibrations in CY3s, which may not be seen 
 clearly  from a single K3 fibered representation  
without using the method given by \cite{hosono1,hosono2}
\footnote{$\sharp$ of K3 fibrations in each CY3 phases means the number of 
divisors, $ J_i$ with $c_2 \cdot  J=24$ under our basis. 
These CY3s have the possibility to have other K3 fibers which can not be seen as a sub dual polyhedron. If one take another base then this number 
may change.}.   
\par
First,  we   use   the  heterotic-type IIA string duality, that is,
if a CY3  admits both 
K3 and $\text{\bf T}^2$ fibrations  with at least one  
section then type IIA string on the CY3
 is dual to  a heterotic string on $\text{K}3\!\times\! \text{\bf T}^2$
\cite{vafa1,vafa2,aspinwall,lerche,witten0,aldazabal1,aldazabal2,candelas1,
berglund,candelas3,bershadsky,lerche0,candelas2}. 
\par
Second, we  use the heterotic-heterotic 
string duality, that is, 
if there are  double K3 fibered CY3s then there are 
two heterotic string compactifications  depending on 
 which K3 fibrations are  used in the compactification\cite{gross}. 
Furthermore, we can extend  to  the  heterotic-heterotic string duality  
between two toric representations with the topologically equivalent 
CY3s. 
\par 
Third, we use the heterotic-type IIB string duality to  
examine how to relate type IIA side to 
the heterotic string sides  in   the strong or weak coupling regions
by calculating discriminants of a mirror  polynomial of CY3 in our case 
and taking an appropriate parameter  limit. 
They correspond to 
the   gauge symmetry  of heterotic string  side which 
comes     from  the  
four contributions:   
(one K3 fibration )$\times $(the other K3 fibration) $\times$ ( d=6 tensor) 
$\times$ (further $ T^2$ compactification). 
\par
By changing a parameter 
 to another parameter in the mirror polynomial, 
we can exchange the weak coupling  
region with a strong coupling region. 
For a toric representation of a CY3 with single K3 fibration phase,  
we can  see  only one-side  contribution of K3 fibrations, i.e.,
weak coupling region only or strong coupling region only. 
By identifying a   double K3 fibered CY3 phase  with  
a single K3 fibered CY3 phase,  we can identify strong coupling  
region physics which is not seen clearly 
with the weak coupling region physics

\par
We examine    two toric representations with the same Hodge numbers,
\newline 
($h^{1,1},h^{1,2})=(5,185)$  
in  two series, (III) of {\bf CFPR} model and (IV) of {\bf HLY} 
model mainly 
\footnote{The definitions of the  models are in the next section.}
     which may  satisfy above three dualities and show their relations.  
Furthermore, $ \Delta n_T \geq $ 3 case is the touchstone 
of the identification of (III) and (IV), since 
the properties of (III) and (IV) are different, i.e.,  
the ADE type singularity appears in (III)
and no ADE type singularity does in (IV). 
The  relations of two toric representations with the 
same Hodge numbers, ($h^{1,1},h^{1,2})=(8,164)$ in (III) and (IV) 
and the heterotic-type IIB 
string dualities of them are in \cite{mmabe}.
\par
One  aim   in  examining   heterotic-type IIB string dualities   
  in   \cite{mmabe} is  to show an interplay of perturbative 
gauge field  and non-perturbative one by using  monodromies and 
discriminants 
\newline  
$\bullet$ For (III) representation with $J_1$ identified with the dilaton
\footnote{There are two parameters, $b_1$ and $b_2$ due to double K3 
fibrations with $b_i =e^{- 2\pi t_i}, c_2\cdot  J_i =24$.(i=1,2).
The explanations of $t_i$ and  $J_i$ are in section 3. 
If  we can identify a parameter in the 
 discriminants in IIB side  with   $t_i $  in type IIA side by mirror map     
then $b_i\rightarrow 0$ 
 is  a weak coupling region and $b_i=1$ will be a strong coupling  region.},
\newline
the gauge symmetry in heterotic string side comes from 
\newline
(I of  K3$_2$ )$\times$(
U(1)$^{\otimes \Delta n_T}$ of  tensor)
 ($A_2$ of  $T^2$ cmpt.) $\leftarrow$ strong
\newline
$\times$ (ADE sing. of  K3$_1$ ) 
$\leftarrow$ weak
\newline
$\bullet$ For (IV) representation with $t_2$ identified with the dilaton,
\newline 
the gauge symmetry in the heterotic string side comes from 
\newline
(I of the generic  K3$_2$ )
 $\times $ (U(1)$^{\otimes \Delta n_T}$ of  tensor)
$\times$ (A$_2$ of  $T^2$ cmpt.) $\leftarrow$ weak 
\newline
$\times$ ( remains of ADE in  K3$_1$   ) $\leftarrow$ strong 
 \newline
(If we take $t_1$ as the dilaton, then the remains of ADE singularity   
appear in the   weak coupling region of $b_1$.) 

The higher derivative couplings of vector multiplets $X$
to  the Weyl multiplet $W$ of conformal N=2 supergravity can be 
expressed as a power series:  
$K(X,W^2)=\Sigma^\infty_{g=0}K_g(X)( W^2)^g$.
Suppose that a holomorphic prepotential of genus zero 
in heterotic string of D=4 side is given by tree and one-loop contributions.
$K_{H}^{(g=0)}=S(TU-\Sigma_i C^i C^i )+ {\cal K}_{H}^{(1)}(T,U,C) 
+{\cal K}_{H}^{(NP)}(e^{-2 \pi S },T,U,C,C^\prime,V)
\in {\cal K}_{H}^{(0)}+{\cal K}_{H}^{(1)} $.
T and U are  two Abelian vector multiplets 
that  contain the Kaluza-Klein gauge bosons 
of the torus and the corresponding toroidal moduli. 
The scalars $C^i$, $i=1, \cdots,$ rank $(G)$   
in a Cartan subalgebra of $G$ are flat directions of the  effective 
potential and at generic values in their field space the gauge 
group is broken to $[U(1)]^{{\text rank}(G)}$.  
A vector multiplt that comes from D=6 tensor multiplet  
and   contains a candidate of  dilaton is denoted by S.
$ {\cal K}_{H}^{(NP)}$ summarizes the 
space-time instanton correction to ${\cal K}_{H}$, i.e.,
 containing sum of  trilogarithmic function,  and ${\cal K}^{(1)}$ is the 
dilaton independent one-loop contribution.  
  In a phase  where  both T and S  have candidate of dilaton  
i.e.,in a double K3 fibered phase, 
if under  $S \leftrightarrow  T$ exchange, 
${\cal K}_{H}^{(NP)} \rightarrow P_3^{(NP)}(T,U,C,C^\prime,V)$ and  
${\cal K}_{H}^{(1)}=P_3^{(1)}(T,U,C^i)+ \cdots$ 
for $S \rightarrow \infty$ then the trace of  heterotic-heterotic duality 
exists where $P_3$ is some triple couplings and   independent on  S. 
$C^\prime$ can be additional vector multiplets or dual tensor-vector 
multiplets  that   
are  of non-perturbative orgin and do not have the canonical 
couplings to one-side  dilaton, S. 
We would like to  seek these phenomena  
 ocurring between two representations.
\footnote{  When  taking a  strong coupling region limit   
such as one of $J_2=0, c_2 \cdot J =24$ 
in a double K3 fibered phase in {\bf CFPR} model then 
we  may find a  single K3 fibered phase in {\bf HLY} model   
that corresponds to this situation.    
 In {\bf CFPR} model side,
$C^{\prime} $ can not be represented by a  toric divisor   however         
$C^{\prime}$ may be   a toric divisor in the single K3 fiberd 
phase of {\bf HLY} model.  (In general, $C^{\prime}$ can be seen in the 
extremal transition \cite{klemm}. However, we would like to relate 
them to the modular forms or the characters.  )
We would like to   
 derive the information for  a compact form 
 of   trilogarithmic function contribution   
and to compare it with that on local CY3 case  because 
    (IV) model side  partition function  
   will  be represented in  a simple form. We would like to express 
  sum of trilogarithmic function as  \cite{saito1,saito2}.}.     

\par 
We also discuss how  to identify two representations.     
Some  identifications of CY3 phases have been 
done by using dual polyhedra
\cite{candelas4,candelas0,kreuzer,skarke}.
The method  given by  \cite{hosono1, hosono2} is powerful to 
 see the property of CY3s and their relations.
There are several works that discuss the relation of 
elliptic fibered CY3s with $\text{\bf F}_0$ base and $\text{\bf F}_2$ 
base  \cite{vafa1,witten2,theisen, curio, mabe}. 
The investigation  in Sec. 4  
is based  on   the topological invariant  calculation 
done by S. Hosono \cite{hosono0} 
and  serves as  an extension  of the earlier works.
The organization of this article is  as follows: 
\newline
\indent
1.  Introduction 
\newline
\indent
2.  Why we compare various models ?
\newline
\indent
3.  The method to identify toric representations 
\newline
\indent
4. The relation among (III), (IV) and (V) models 
\newline
\indent
5. Future problem
\section{Why  we compare various representations  ?}  
There are several series of CY3s   that  
 arises naturally from the heterotic-type IIA string duality. 
The starting point is  $E_8 \times E_8 $  heterotic string compactified 
on $\text{K}3\!\times\!\text{\bf T}^2$ with $G_1\!\times\!G_2$ bundles  
with instanton numbers  ( $k_1$, $k_2$) such that    $ k_1\!+\!k_2\!=\!24$
\cite{aldazabal1,aldazabal2}   
\footnote{$(k_1\!=\!12\!+\!n^0,\quad k_2\!=\!12\!-\!n^0) $
where $n^0$ is introduced for convenience.
$G_1$ and $G_2$  come from each $E_8$. }. 
Using the index theorem  and anomaly cancellation  
condition, we find spectra of D=4 N=2 heterotic string vacua, 
which are  related  to the Hodge numbers of CY3-folds in typeIIA string 
side with the same spectra. We list  four   series that have  dual,  type IIA  string  on CY3s
\cite{aldazabal1,aldazabal2}\footnote{   Furthermore, there are 
three versions of these  series 
 by changing the type   of elliptic fiber. 
 A-chain version is in the  \cite{aldazabal2,candelas3},
where the elliptic fiber of the CY3 
is $\text{\bf P}(1,2,3)[6]$.   
The extension  to B or  C versions  
with elliptic fiber $\text{\bf P}(1,1,2)[4]$ or $\text{\bf P}(1,1,1)[3]$ 
are also possible.}
(CY3s  used in (I) and (II) series are in tables 1,2 and 3. )
\begin{enumerate}
\item 
In the first series,  
\newline
$E_8 \rightarrow G_1 =I$
\newline  
$E_8 \rightarrow G_2= \{I, A_1,A_2, D_4, E_6, E_7, E_8      \}$ 
 that depends on $n^0$.   
\newline
We call this series as (I) (terminal case). ( see table 2)
\item 
The second series,
\newline
$E_8 \rightarrow G_1 \not= I,i.e., G_1=\{A_1,A_2,A_3, \cdots \}$
\newline  
$E_8 \rightarrow G_2= \{I, A_1,A_2, D_4, E_6, E_7, E_8      \}$ 
 that depends on $n^0$.   
\newline
We call  this series as (II).  
\end{enumerate}
We follow the notations, such as (I), (II), etc. given in  
in those of   \cite{mmabe,mabe}.  
\newline
\indent 
Most of  CY3s in (I) and (II)   can be extended to be   
CY3s   with extra blow ups    by adding appropriate toric points 
\cite{aldazabal2,candelas0}. ( (III) and (IV) are in tables 4 and 5)
They are classified to   third (i.e., an kind of (I)$^\dagger$) 
or  forth  series (i.e., an kind of (II)$^\dagger$ 
\footnote{(I)$^\dag$/(II)$^\dag$ means the modified (I)/(II) in 
\cite {candelas1} with extra tensor multiplets.}). 
(III) of {\bf CFPR} model   is a  similar  extension from (I)
\footnote{ The difference between  the 
dual polyhedron of  (I)$^\dag $ 
and (III)  is as follows \cite{candelas1,candelas2,candelas3}.  
The dual polyhedra of case  (I)$^\dag$ have the modified 
 dual polyhedra of  K3 part. 
For the case  (III) in the A series,  the dual polyhedron of 
K3 part    is  not modified.  
The highest point in  the  
additional  points is  represented by the  weights of the 
K3 part of the terminal A series in this base.
One   point such  as  $(0,\ast,\ast,\ast)$   is also  
represented by   the part of the  weight of the terminal K3 part. 
The following element in SL(4,{\bf Z}) can 
 transform   these polyhedra  
into the dual polyhedron given by   \cite{candelas2}.
$
\left(
\begin{array}{cccc}
1 & 0&1&2 \\
0&1& 2& 3 \\ 
0&0 & 1 &2\\
0&0 &  -1& -1\\
\end{array}
\right).
$
In the base of \cite{candelas2},
the additional points make a line with $x_4=-1$. }
and (XI) of  {\bf CFPR} model  from (II).   
We see  that (I$^{\dagger}$)  and  (III) are equivalent 
\footnote{
Their triangulations coincide by identifying the replaced vertices. 
Furthermore, U(1) charges defined in appendix 1  for  (III) 
of  \cite{candelas2} matches with the that  for    (I)$^\dag$  of  
\cite{candelas3}.}.
\par
There is a  quite different type extension of CY3s 
which is  given by \cite{yau}. 
Keeping the  K3 fibration with the same weight,  
a weight of base $P^1(1,s)$  is changing
 in this representation of CY3s.  
We call them (IV) and (VI) of  {\bf HLY } model, 
which may  relate to be (I)$^\dagger$ and  (II)$^\dagger$
of {\bf CFPR} model 
\footnote{The relation of   toric realizations of (II)$^\dagger$  and (XI) 
is similar to that of (I)$^\dagger$  and (III). They will be 
coincide with each other. }.  
\newline  
There is another type  of the extension that has  
a triple K3 fibration at most \cite{theisen}. 
We call this (V) of {\bf LSTY} model. 
\begin{quotation}
3.  The third series, 
\newline
$E_8 \rightarrow G_1 =I$
\newline  
$E_8 \rightarrow G_2= \{I, A_1,A_2, D_4, E_6, E_7, E_8      \}
\times U(1)^{\otimes \Delta n_T}$ 
 due to additional tensor multiplets.   
We call this series as (I)$^\dagger$  or  (III) of {\bf CFPR} model.  
(V) of {\bf LSTY} model are    also in this series.  
(IV) of {\bf HLY} model may relate  or  belong to this series.
(see table 3 and 4)

4.  The forth series, 
\newline
$E_8 \rightarrow G_1 \not= I, G_1=\{A_1,A_2,A_3, \cdots \}$
\newline  
$E_8 \rightarrow G_2= \{I, A_1,A_2, D_4, E_6, E_7, E_8      \}
\times U(1)^{\otimes \Delta n_T}$. 
 (II)$^\dagger$  or  (XI), 
  extensions  from  {\bf CFPR} model  are   in this series. 
(VI) of  {\bf HLY} model    may  
relate or  belong to  this series. ( see table 7)
\end{quotation}  
Each CY3     can be realized by a hypersurface 
in a  toric variety.
A dual polyhedron of them    contains 
some  sub dual polyhedra of K3 part,  which   
 can be  K3 fibrations in some phases.    
 Base surfaces under  the elliptic fibration of these CY3s are  
 blow-ups of the $a$th Hirzebruch surface, $\text{Bl}(\text{\bf F})_a$. 
We give some explanations of models and   
list  four dual polyhedra with $(h^{1,1},h^{2,1}) \!=\!(5, 185)$  
that we deal with in this paper.
\newline   
\noindent
(\text{\bf I} and \text{\bf II}) 
{\tt Models   of  Aldazabal, Ibanez, Font, Quevedo and Uranga}
 \cite{aldazabal1, aldazabal2} ({\bf AIFQU} model) 
\newline
A dual polyhedron of this representation in \cite{candelas1}    
 contains    two    sub dual polyhedra  of    K3 part  as 
$(0,\ast,\ast,\ast)$ or $(\ast,0,\ast,\ast)$.   
One of which     varies  according to  the instanton numbers.  
CY3s   have  a $\text{\bf P}^1 (1,1)$ base under this  K3 fibrations,
(see table 1).
CY3s  have a $F_{n_0}$ base under a  elltiptic fiber. 
\newline  
\newline
\noindent
 $(\text{\bf I}^{\dagger})$
{\tt Models of Candelas, Font, Perevalov and Rajesh} 
\cite{candelas1,candelas3} ({\bf CFPR} model)
a   dual polyhedron  in \cite{candelas3}of  {\bf CFPR} model 
\begin{align*}
{} &{} \phantom{\longleftarrow\quad } \text{difference from (III)}  \\
(0)\   (\ph 0,  \ph 0,  \ph 0, \ph 0)\  &{}   \\
(1)\  (\ph 0,  \ph 0,  -1,  \ph 0)\  &{} \\
(2)\  (\ph 1,  \ph 0,  \ph 2,  \ph 3)\  & {} \\
(3)\  (\ph 1,  \ph 2,  \ph 2,  \ph 3)  
\ &\longleftarrow\  (\ph 1, \ph 2, \ph 6, \ph 9 )^{(\III)}\\
(4)\   (\ph 0,  \ph 0,  \ph 0,  -1)\  &{}    \\
(5)\  (-1,  \ph 0,  \ph 2,  \ph 3)\  &{}\\
(6)\ (\ph 0,  -1, \ph 2,  \ph 3)  
\ &\longleftarrow\   (\ph 0, -1, \ph 0, \ph 0 )^{ (\III)} \\
(7)\  (\ph 0,  \ph 1,  \ph 2,  \ph 3)  
\ &\longleftarrow\  (\ph 0, \ph 1, \ph 4, \ph 6 )^{(\III)} \\
(8)\  (\ph 1,  \ph 1,  \ph 2,  \ph 3)  
\ &\longleftarrow\   (\ph 1, \ph 1, \ph 4,\ph  6 )^{(\III)} \\
( 9) \ (\ph 0,  \ph 0,  \ph 2,  \ph 3) \ &{}.       
\end{align*}
$(\I^{\dagger})$ have extra one  cones of  (1)(1,1,2,3) and (8)(1,1,2,3) 
in addition to those of  (I). K3 part is the same as  those in (I) and (III).
\newline
(\text{\bf III}) {\tt Models   of  Candelas, Perevalov and Rajesh}
 \cite{candelas2} ({\bf CFPR} model) 
\newline
 a  dual polyhedron  in  \cite{candelas2} of {\bf CFPR} model 
(We use the right hand side   in this paper.)
{\allowdisplaybreaks      
\begin{align*}
{}  \text{SL}(4;\text{\bf Z})\ & \text{trans.} {}\\
(0)\quad  (\ph 0,  \ph 0,  \ph 0,  \ph 0) 
\rightarrow&\   (\ph 0, \ph 0, \ph 0, \ph 0),\\
(1)\quad (\ph 0,  \ph 0,  \ph 1,  \ph 2) 
\rightarrow&\  (\ph 0, \ph 0,  -1, \ph 0),\\
(2)\quad  (\ph 1,  \ph 0,  \ph 0,  -1) 
\rightarrow&\  (\ph 1, \ph 0, \ph 2, \ph 3),\\
(3)\quad (\ph 1, \ph 2,  \ph 0,  -1)  
\rightarrow&\   (\ph 1, \ph 2, \ph 6, \ph 9),\\
(4)\quad (\ph 0,  \ph 0,  -1,  -1) 
\rightarrow&\   (\ph 0, \ph 0, \ph 0, -1),\\
(5)\quad  (-1,  \ph 0 , \ph 2,  -1) 
\rightarrow&\   (-1, \ph 0, \ph 2, \ph 3),\\
(6)\quad  (\ph 0,  -1,  \ph 1,  -1) 
\rightarrow&\   (\ph 0, -1, \ph 0, \ph 0),\\
(7)\quad  (\ph 0,  \ph 1 , \ph 1,  -1) 
\rightarrow&\  (\ph 0, \ph 1, \ph 4, \ph 6),\\
(8)\quad  (\ph 1,  \ph 1,  \ph 0,  -1) 
\rightarrow&\  (\ph 1, \ph 1, \ph 4, \ph  6),\\
(9)\quad  (\ph 0,  \ph 0,  \ph 1,  -1) 
\rightarrow&\  (\ph 0, \ph 0, \ph 2,  \ph 3). 
\end{align*}}
A dual polyhedron  contains    two   
  sub dual polyhedra  of    K3 parts: $(0,\ast,\ast,\ast)$ 
and $(\ast,0,\ast,\ast,\ast)$.  
One of which    varies  according to  the number  
of tensor multiplets.  CY3s have        
 a $\text{\bf P}^1 (1,1)$ base under this  K3 fibrations.
The latter K3 part   is always   $\text{\bf P}^3(1,1,4,6)[12]$. 
 CY3s   contain  some  double  K3 fibration phases.  
 The $h^{1,1}\!=\!5$ case  has  $\Delta n_T\! =\!2$
and  $\text{\bf P}^3(1,1,4,6)[12]$ as both K3 fibrations.
\newline
\noindent
(\text{\bf IV}) {\tt Models  of  Hosono, Lian  and Yau }  \cite{yau}
 ({\bf HLY } model) 
\newline
CY3s  are realized by the weighted projective hypersurfaces 
$
\text{\bf P}^4(1,s,s+1,4s+4,6s+6)[ 12s+12]
$
with  $\{ s\!=\!1,2,3,4,6,8,12 \}$.
CY3s have  single   K3 fibration phase  with fiber  
$\text{\bf P}^3(1,1,4,6)[12]$ as $(0,\ast,\ast,\ast)$ and 
 base, $\text{\bf P}(1,s)$.   
$h^{1,1}\!=\!5$ case has  $s\!=\!\Delta n_T\!=\!2$.
\par
Using toric data, we can  examine the heterotic-type IIA 
duality.
The heterotic-type IIA string duality for (III)
and (I$^\dagger$)  has been made 
clear in \cite{candelas3,candelas4}, which we review  at first. 
The differences between the Hodge numbers of (I) and (III)
or (I$^\dagger$) with
$k_1+k_2+\Delta n_T=24,k_1=12+n^0-\Delta n_T$,
$n^0=\Delta n_T$ are given by    
\begin{eqnarray}
\Delta h^{2,1}&=& -h^{2,1} \mid_{\text{in (I)}}
+h^{2,1} \mid_{\text{in (III)}}=-29 \Delta h^{1,1},\nn  
\\
\Delta h^{1,1}&=& -h^{1,1} \mid_{\text{in (I)}}
+h^{1,1} \mid_{\text{in (III)}}
=\Delta n_T.\nn
\end{eqnarray}
\noindent
 a   dual polyhedron in   \cite{yau} of {\bf HLY } model
\begin{align*}
{} &\phantom{\longrightarrow\quad}
\text{difference  from (III)} \\
(0)\ (\ph 0,  \ph 0,   \ph 0,   \ph 0)\ &{} \\
(1)\    (\ph 0,  \ph 0,  \ph 0,    -1)\ &{} \\
(2)\    (\ph 0,   \ph 0,  -1,  \ph 0)\ &{} \\
(3)\     (\ph 0,   -1,  \ph 0,   \ph 0)\ &{} \\
(4)\     (-1,\ph 0,   \ph 0,  \ph 0)
\ &\longleftarrow\
(-1,  \ph 0,  \ph 2,  \ph 3 )^{(\III)} \\
(5)\   (\ph 2,  \ph 3, \ 12,  \ 18)
\ &\longleftarrow\
(\ph 1,  \ph 0,  \ph 2,  \ph 3 )^{(\III)}\\
(6)\   (\ph 1,  \ph 2,   \ph 8, \  12)
\ &\longleftarrow\
( \ph 1,  \ph 2,  \ph 6,  \ph 9 )^{(\III)} \\
(7)\  (\ph 0,  \ph 1,  \ph 4,   \ph 6) \ &{} \\  
(8)\  (\ph 1,  \ph 1,   \ph 6,   \ph 9)
\ &\longleftarrow\
(\ph 1,  \ph 1,  \ph 4  ,\ph 6 )^{(\III)}\\
(9)\  (\ph 0,  \ph 0,   \ph 2,   \ph 3) \ &{}        
\end{align*}
The number of the tensor multiplets 
$n_T$  in (IV) is given by 
$
n_T=h^{1,1}\left(\text{Bl}(\text{\bf F}_2)\right)-1
=d_1-2d_0-1
=s+1, \ (s\geq 2),
$
where $d_i$  denotes  the number of 
$i$-dimensional cones of the fan that  describes 
the base $\text{Bl}(\text{\bf F}_2)$. 
The Hodge numbers and $n_T$ in (IV) and (V) coincide   
with those in (III).
It seems that there exists a heterotic string on 
$\text{K}3\!\times\!\text{\bf T}^2$ that is  
 dual to both the   type IIA string 
compactified  on  CY3 of (IV) and 
that on a CY3 of (V).
\newline
Another candidate of single K3 fibered CY3 representation is 
\newline
\noindent
{\bf (VI)}
{\tt Models  of  Hosono, Lian  and Yau }  \cite{yau}
 ({\bf HLY } model) 
\newline
$\text{\bf P}^2(1,s)$ based 
${\text{\bf P}}^3(1,1,3,5)[10]$ fibered CY3s, 
$
\text{\bf P}^4(1,s,s+1,3s+3,5s+5)[ 10s+10]
$
with  $\{ s\!=\!1,2,3,5,7,9,10  \}$.
This representation 
relates to the forth series of heterotic type IIA 
string duality   with $ G_1=A_1$  in A series.  
Some of  CY3s    satisfy the following anomaly 
free conditions for $s=\{2,5,7,9\}$\cite{mmabe}  
\footnote 
{$n^0=0$ case with ${\text{\bf F}_0}$ based CY3 and  $n^0=2$ case 
with ${\text{\bf F}_2}$ based CY3  in (II) have the 
different Hodge numbers.  
Therefore, for some CY3s with small s, the anomaly free-conditions are 
changed.}.  As the result of the comparison of 
 (VI)  and (II) in A series with $G_1= A_1$, we obtain 
\newline
\indent$\Delta h^{2,1}=h^{2,1}\mid_{\text{(VI)}}- h^{2,1} \mid_{\text{(II)}}$, 
~$\Delta h^{1,1}=h^{1,1}\mid_{\text{(VI)}}- h^{1,1} \mid_{\text{(II)}}
= \Delta n_T$, 
\newline
\indent
$h^{2,1}|_{{\text{(II) with}} A_1}
=h^{2,1}|_{\text{(I)}} -(12n+29)$, 
$\Delta h^{2,1}= -(29-12) \Delta h^{1,1}= -17 \Delta h^{1,1}$. 
with 
$k_1+k_2+\Delta n_T=24,k_1=12+n^0-\Delta n_T$,
$n^0=\Delta n_T$.
\newline
For  $K3 \times T^2$ side,  $12n+29$   is calculated  
by the index theorem and denotes 
the number of $G_1$  charged hyper multiplet 
fields\cite{candelas0,aldazabal2}. 
We substitute  $n=n^0-\Delta n_T$ for the extra tensor 
multiplets case instead $n=n^0$.   
Similar extensions to other G$_1 \not= I $ gauge group  
in A chain, B and C chain versions also seem to be possible \cite{mmabe}. 
We also suppose the existence of the double K3 fibered CY3s, which  
is denoted as  (IX).
They  can be obtained by the extension  
from   A series with $G_1=A_1$ in model (II). 
\newline
\noindent
(\text{\bf V}) {\tt Model  of Louis et al.} \cite{theisen} 
 ({\bf LSTY} model )
\newline   
a  dual polyhedron  in  \cite{theisen} of {\bf LSTY} model 
(It can be obtained by the modifications of (III) or $(\I^{\dagger})$)
{\allowdisplaybreaks
\begin{align*}
(0)\ (\ph 0,  \ph 0,  \ph 0,  \ph 0)\ &{}\\               
(1)\ (\ph 1,  \ph 1,  \ph 2,  \ph 3) 
\  &\longleftarrow\
 (\ph 1, \ph 1, \ph 4, \ph 6) ^{(\III)} \\ 
(2)\  (\ph 1,  \ph 0,  \ph 2,  \ph 3)\ &{} \\                   
(3)\ (\ph 0,  -1,  \ph 2,  \ph 3)
\   &\longleftarrow\
 (\ph 0, -1, \ph 0, \ph 0)^{(\III)}\\ 
(4)\ (-1, -1, \ph 2, \ph 3)
\    &\longleftarrow\
(\ph 1,  \ph 2,  \ph 6, \ph 9)^{(\III)}\\ 
(5)\ (  -1,  \ph 0,  \ph 2, \ph 3)\ &{} \\       
(6)\ (\ph 0,  \ph 1,  \ph 2, \ph 3)
\  & \longleftarrow\
 (\ph 0, \ph 1, \ph 4, \ph 6)^{(\III)}\\ 
(7)\  (\ph 0,  \ph 0,  \ph 2,  \ph 3)
\ &\longleftarrow\  
\text{the point absent in ``17''   {\cite[p.20]{theisen}}}
\\
(8)\ (  \ph 0,  \ph 0,  -1,  \ph 0)\ &{} \\        
(9)\ (  \ph 0,  \ph 0, \ph  0,  -1)\ &{}       
\end{align*}}
A toric representation of (V)  with $h^{1,1}\!=\!5$
contains  three dual sub polyhedra of K3 part. 
All K3 part  are realized by 
$\text{\bf P}^3(1,1,4,6)[12]$.
\footnote{ The dual polyhedron of (V) 
coincides with that of   $(\I^{\dagger})$ except one vertex:
$(1, 2, 2, 3)^{(\III)} \rightarrow (-1,-1,2,3)^{(\V)}$. 
By this, the  existence of three symmetric K3 sub dual polyhedra  
can be seen apparently:   
$
 \{(5)(-1,0,2,3),\ (\ast,0,\ast,\ast),(2)(1,0,2,3)\}\ 
\leftrightarrow
\{(3)(0,-1,2,3),\ (0,\ast,\ast,\ast),(6)(0,1,2,3)\}\  
\leftrightarrow
\{(4)(-1,-1,2,3),\ (\ast,\ast,\ast,\ast),(1)(1,1,2,3)\}\ . 
$
}
By  these toric data, we can see the structure of the 
K\"ahler moduli spaces CY3\footnote{We follow the result, notations and 
definitions of \cite{hosono1,hosono2}.}. 
Given a singular ambient space, we have in general many 
phases in the associated K\"ahler moduli space of the nonsingular 
ambient space. 
\section{The method to identify toric representation}
We discuss the case when 
 two dual polyhedra  have  no  
twisted sectors ( = non-toric degree of freedom).
The method that we use is given by \cite{hosono1,hosono2}, that is, 
to derive Gromov-Witten invariants, to compare them directly 
and to examine the relations of CY3 phases. 
It is the most effective and rigorous way.  
Especially, if some Gromov-Witten invariants of  CY3s are  those   of
Del Pezzo surfaces, the comparison is very easy.
By fixing  U(1) charges, Q such as in appendix 1, 
we first have to calculate Mori-cones\cite{mori} 
and K\"ahler cones in each phase\footnote{  They  satisfy 
$\{  Q_1,\cdots,  Q_9 \}=\{  J_1,\cdots, J_5 \}
\cdot  
\begin{pmatrix}
& \ell_1\\
& \ell_2\\
& \ell_3\\
& \ell_4\\
& \ell_5\\
\end{pmatrix} 
$.
By matrix notation, 
$ J_{j}$ and $ Q_{i} $ are column vectors and 
$ \ell_i$ are  row vectors. $i=1,\cdots 5 for h^{1,1}=5$ case. 9 denotes the 
number of the points in the dual polyhedra of a CY3-fold. }.
Mori-cones generated by the holomorphic curves  $\{ \ell_j\}$
are the  dual  of  K\"ahler cones generated by 
$\{ J_i\}$  \footnote
{The complexified K\"ahler class $J$  is givem by 
$ J=\sum^{h^{1,1}}_{i=1}t_i  J_i
 \in H^2(M;\text{\bf C})$.} 
\footnote{
Both $ \ell_j$ and $ J_i$ are represented 
by the common  dual basis in each models, $ m_i $ and $ D_i$ 
such as  $ J_{j}=\Sigma  D_{i} A_{ij}     $ 
\footnote{In general, $ c_2 \cdot  J_j \not= 
\Sigma A_{ij} c_2 \cdot  D_i $.} and   
 $ \ell_{i}=\Sigma S_{ij}  m_{j } $. 
$ D_{i} $ are  toric divisors    
 corresponding  $ Q_i$ charges.  
$ D_i \cdot  m_ j=\delta_{ij}$. 
$A_{ij}$ and $S_{ij}$ are 5 $\times$ 5 
transformation matrices.} 
\footnote{
We can see that the volume of the curve 
$ \ell=\sum_{j}  n_{j}    \ell_j$  
measured by  $J$ is    
$\text{vol}_{ J(  \ell)}=\sum^{h^{1,1}}_{i=1} n_i t_i$. 
}.
The Gromov-Witten invariants $N(\{n_i\})$  are defiend by
 the instanton corrected  Yukawa coupling $K_{x_i,x_j,x_k}$. 
 $N(\{n_i\})$ is the instanton number of the rational curves $C$
of  multidegree $\{n_i\!=\!\int_C J_i\}$.
The algebraic coordinates 
$\{x_i\}$ and  the special coordinates $\{t_i\}$
are related to  the mirror map via Mori-vectors \cite{hosono1,hosono2}. 
{\small
\begin{align*}
&K_{t_i,t_j,t_k}(t){}_{\text  {II}}^{g=0}
={1 \over w_0(x(t))^2 }
\sum_{lmn} 
{\partial x_l \over \partial t_i}
{\partial x_m \over \partial t_j}
{\partial x_n \over \partial t_k}
{\cal K}_{x_l, x_m, x_n}(x(t)),
\nonumber \\
&={\cal K}^{0}_{ijk}+ 
\sum_{\{n_l\}}
 N(\{ n_l\} )n_i n_j n_k\,  
\frac{\prod_l q_l^{n_l}}{1\!-\!\prod_l q_l^{n_l}},
\end{align*}
}

where $q_i\!=\!\text{e}^{-2\pi  t_i}$. 
Integrating back yields a trilogarithmic function. 
${\cal K}^{0}_{ijk}$ is the classical part
of the Yukawa couplings
\footnote{
$
{\cal K}^0_{ijk}:=  d_{ijk} t_i t_j t_k,
\quad 
d_{ijk}=\int_{\text CY3}  J_i \wedge  J_j \wedge  J_k$.
To get the Yukawa coupling from this  notation, 
some additional  normalizations  factors
are necessary: $t_i^3 \rightarrow \frac{1}{3!} t_i^3$ 
and  $t_i^2 \rightarrow \frac{1}{2!} t_i^3$. }.
The dual polyhedra that we compare 
do not coincide with each other  by  SL(4,\text{\bf Z}) transformation.
Nevertheless, in some phases, topological invariants 
happen to  match. Note that Mori vectors do not 
match even if they are equivalent CY3 phases 
between the different models. 
We use the following theorem and  the  sub steps.
Wall's Theorem says that   
 the agreements of classical invariants,
 $c_2 \cdot   J$ and ${\cal K}^0$,  lead to 
the  agreements  of  topology as well as Gromov--Witten invariants 
\cite{wall} in two CY3s.
We can narrow down   candidates of the transformation 
matrix  by comparing values of $c_2\cdot  J_i$.
\begin{itemize}
\item
criterion 1 : 
\newline
If $c_2 \cdot  J $  and   $ {\cal K}^0$ 
match  then it leads to the agreement of 
the Gromov--Witten invariants,  
$ N (\{n_i\})$. 
In this case,  the number of the K3 fibrations in  two phases is the same
\footnote{
In these cases, we can see some mappings of the ambient space data 
between two equivalent  whole K\"ahler cones  of two models.}.  
\item 
criterion 2 : 
\newline
To compare the two phases  ( regardless of  their
jurisdiction, i.e.,    the different models or 
 the same models ), we can  use a candidate of transformation 
matrix of topological 
invariants by combining some transformation matrices of divisors. 
We can make such a matrix by replacing a divisor of the 
original phase by another divisor.
 These  divisors  have   the same 
$c_2 \cdot  J_i$ and $d_{ijk}$.   
If this matrix is integer valued and 
transforms topological invariants, then 
these two phases are  the same ;
Let $( J_i,  \ell_j)$ and 
$({ J'}_i,{ \ell'}_j)$ be some generators of the  
K\"ahler  and the Mori cones of the two equivalent  CY3 phases. 
An integer-valued matrix of divisors  such as  
$
{J'}_i^{(B)}=\sum_{j=1}^{h^{1,1}} J_j^{(A)}
( M_{(AB)}){}_{j,i} ,
\quad
 {\ell'}_i^{(B)}=\sum_{j=1}^{h^{1,1}}
( M_{(AB)})^{T}{}^{(-1)} {}_{i,j}  \ell_j^{(A)},
$  
transform topological invariants,
$c_2 \cdot  J_j'= \sum_{j=1}^{h^{1,1}}   
M_{\text{(AB)}}{}_{ij}   
c_2 \cdot    J_i'$,
$d_{ijk}'=\int_{\text{ CY3}} 
 J_i'\wedge   J_j' \wedge  J_k'
=\sum_{lmn}  M_{\text{(AB)}}{}_{il}
   M_{\text{(AB)}}{}_{jm}   M_{\text{(AB)}}{}_{kn}
\int_{\text {CY3}}  
  J_l \wedge  J_m \wedge  J_n,$
and 
$
N(\{n_i^{(A)}\})=N'(\{ {n'}_i^{(B)}\}),
\quad 
\text{for }\  {n'}_i^{(B)}=\sum_{j=1}^{h^{1,1}}n_j^{(A)} 
( M_{\text{(AB)}}){}_{j,i}$ 
\footnote{A curve $[C]$ admits the expansion
$
[C]=\sum_{i=1}^{h^{1,1}}n_i\ell_i
=\sum_{j=1}^{h^{1,1}}{n'}_j{\ell'}_j.$}
\footnote{
A transformation matrix, $ M$ can 
contain some negative 
integers even if they are in the same phase. 
However,$\{n_i\}$  and  $\{n'_i\}$ are bijective and  should contain 
only positive integers.} 
\footnote{A  transformation matrix 
of  whole K\"ahler cones such as 
$ M_{\text{(AB)}}^{\rm (IV),T} =
A_{\text{(B)}}^{\rm (IV),T}A_{\text{(A)}}^{\rm (IV),T}{}^{-1} $
in the same model can not transform topological invariants, because 
they depend on the ambient space data specified by the triangulation. 
(A and B   denote phase names for example. )
However, modifying or combining   $A$, 
we can get a transformation matrix of  equivalent phases.
Though in one model, all the divisors can not always be represented 
by the data of the ambient space,  they will be transformed or related    
to  those  of the  another model. We confirmed this by  including (V).}.
\end{itemize}
\section{ The relation among (III), (IV) and (V) models}
 There are two characteristic points  about the triangulations  of  
(III), (IV) and (V) with $(h^{1,1},h^{2,1})\!=\!(5,185)$.  

The first point is about the feature of 
the Gromov-Witten invariants.  
All phases of them contain  Del Pezzo 4-cycles, $B_8$
\footnote{$B_8$ is  given by       
$E_8={\bf P}^2(1,2,3)[6]$  fibered 4-cycle and has eight blow up points.  
The property of  Gromov-Witten invariants in $B_8$ 
is ruled by this elliptic fiber.  } in many ways\footnote{
 Most phases have $B_6$ in six ways. For example, 
we can reduce 
 Mori vectors of CY3 in triple K3 fibration phase, 
$\alpha_{10}$ to those of  
 Del Pezzo $B_8$ in six ways.  }.   
The existence    of   $B_8$  in  ${\bf F}_1$ based 
elliptic fibered CY3 with $(h^{1,1},h^{2,1})\!=\!(4,214)$ has  
already been investigated \cite{lerche}.  
$(h^{1,1},h^{2,1})\!=\!(5,185)$ case has one more extra blow up point 
than $(h^{1,1},h^{2,1})\!=\!(4,214)$ case. 

The second point  is that the different triangulations of them
 do not lead to the different phase. 
In general,  a  CY3  phase is specified by a particular 
triangulation of the polyhedron. 
However, some different triangulations ( called phases in this article) 
do not  lead to different CY3 phase \cite{berglund}.   
In that case, the conclusion is that  the singularities  on the submanifolds
  blown up to specify each phase   do not contribute to CY3 phases.   
This property  depends on the  dimensions of the submanifold 
that contains them \footnote{In this case, 
the dimension of the submanifold is codim 2 of CY3 +1.}.  
Some  triangulations in (III), (IV) and (V) resulted into this case. 
\par
There are five phases in {\bf HLY } model, eight phases  
in {\bf CFPR} model and eighteen phases in {\bf LSTY} model  
which are specified by the triangulation ( see table 7). 
The identifications of CY3 phases  
 by the criterion 1 is shown in table 9.
The phases in the same line in table 9 have the same 
topological invariants. Four phases of  a single K3 fibration in 
{\bf HLY } model can be identified with 
four phases of  a single K3  fibration  in {\bf CFPR} model. 
15 phases of (V)  in {\bf CFPR} model can be identified 
with four phases of (III) of {\bf CFPR} model.  
\par
By using criterion 2, 
we can identify  one pair of phases such as
the phase A with a single fibration  of {\bf HLY} model and 
the phase g with a double K3 fibration of {\bf CFPR} model.
We can also identify the
left four  phases of (III) in {\bf CFPR} model   including phases with   
 double K3 fibration  and  the left three phases  in  (V)
 of {\bf LSTY} model including a triple K3 fibration case  
with the phases in (IV) of {\bf HLY} model.  
The result is in table 10.
\par
In conclusion,  there are only five  topologically 
nonequivalent  phases   defined  by the different  
triangulation  in (IV) of {\bf HLY } model     
for $\Delta n_T=2$. 
 The other phases in (III) of {\bf CFPR} model and (V) of {\bf LSTY} model  
 are  equivalent to these five phases. 
Each model that contains topologically equivalent phases is a local  
coordinate representation  of the same  CY3-fold   
\footnote{All   phases in  (IV)  are       
represented by the  simplicial cones. 
However,  half of  (III) phases  and most  of  (V) 
phases are not      simplicial cones.
It is difficult to take five true phases by taking the 
union of  K\"ahler  cones of the equivalent phases 
among (IV), (III) and (V). Because, we must  
 get  all   virtual Mori-vector of the one model  side,   
which corresponding to Mori-vectors of the  equivalent phase 
of the other model  to take the 
intersection. For example, we can not  get the cap of
Mori-vectors of  phase A and g and   $\alpha_{10}$ in (III), 
(IV) or (V) side .}. 
 
\section{Future problem}
In this paper,   we derived the relationship of three CY3 models with 
$(h^{1,1} h^{2,1})\!=\!(5, 185)$.  
We come to the conclusion that  three models and their   extensions  
 satisfy  N=2 and N=1 string dualities\cite{curio1} 
because they are all local representations of the same CY3.       
This is   the starting point of the comparison of (III) of {\bf CFPR} model 
and (IV) of {\bf HLY} model to derive the example of  
 N=2 and N=1 string dualities 
on two toric representations of  the same CY3 model. 
In  $h^{1,1}=5$ case, both  K3 fibrations are the same K3 
  therefore  interplay of nonperturbative and perturbative 
gauge fields, i.e., the trace of  heterotic-heterotic string  duality   
is  not seen \cite{theisen}\footnote{
For   triple  K3 fibered phase of  $\alpha_{10}$=   
phase 17 of  \cite{theisen},  three $t_i$ with $C_2 \cdot J_i=24 $ 
are symmetric in ${\cal K}^0$ therefore, 
the change $S\leftrightarrow T$ is symmetric.    
The IIA string side topological part of prepotential is    
 $ {\cal K}^0=STU-\frac{1}{2}U V_Y^2 -\frac{1}{2}U W_Y+\frac{1}{4}U^3$ by 
changing $t_{i}$ into heterotic side variables.  
$V_Y$ and $W_Y$ are two tensor multiplets \cite{theisen}. }. 
However, this CY3 is  an extension of phase 6 with (4,214) in \cite{theisen}
 which is related to the non-critical string model\cite{Lerche1} 
 though  the part of CY3 where $B_8$ exists  is different.  
We conjecure that the relation of   {\bf HLY} and {\bf CFPR}  models 
 with higher Hodge numbers  
may also  be interpreted by shrinking $B_8$ and flops on    
 (${\bf P}^1(1,1)$ based ${\bf P}^1(1,s)$ fibered ) Hirzebruch surface based  
 elliptic fiberd CY3-fold 
   \footnote{
Type IIB string  side on  (IV) case is  a  theory  on      
 ${\bf P}^1(1,s)$  based ALE fibered one,  which  might  correspond  to  
 the non-perturbative property of (III). }.  
For higher Hodge numbers case, the type of double K3 fibration is 
different and tree level topological three-point function 
 with a dilaton, T  should come from non-perturbative terms of 
 with another dilaton, S  side in the 
double K3 fibration phase. Therefore,     
the  trace of heterotic-heterotic duality will be seen apparently. 

4D $N$\!=\!2 super YM theory  can  also be analyzed as  the 
heterotic strings compactified on 
$K3 \!\times\! \text{\bf T}^2$ in the weak coupling region. 
The threshold correction of case (I) in heterotic string side
 has been given   by the calculation of 
the partition function \cite{moore,kawai, curio2, eguchi1}. 
 By combining their method  with generalized modular forms \cite{Eichler}
  and   the result of local $B_8$ string model \cite{Lerche1,mohri}, 
we will be able to  derive the perturbative  Yukawa couplings  
for the  extra tensor multiplets case by comparing the Gromov-Witten  
invariant data of type II  
\footnote{Take for a simple  example,   
phase 6 of  (4,214) case in\cite{theisen},  
 Mori-vectors of CY3 reduce to those of   Del Pezzo $B_8$ surfaces  
in three ways,  which agrees with  table 8.      By decomposing  
the result     of  \cite{mohri} and replacing some part  to match
 with  brown up $F_2$ data in table 3 of \cite{chiang}, 
we will be able to derive 
the partition function. (See appendix 6 and  \cite{mmmabe}).}  
\footnote{
An  examination   of relations among $h^{1,1}=4$ and $h^{1,1}=5$ models 
and their partition functions    in type II-heterotic 
string web in table  \cite{theisen, curio}
    will be interesting by applying a method in this paper.   
  The phase 5 given by in \cite{theisen} 
  has non-symmetrical  double K3 fibrations and the trace of 
 heterotic-heterotic string duality, which is contained in  
CFPR model with (8,164).      }.      
    
Furthermore, we would like  to  compare the 
partition function on global CY3  
and  on local CY3 \cite{losev} 
\footnote{For example, the  base of CY3-fold of phase 6 with (4,214) is 
$F_2$ with one point blown up  and the same  as those of  
CY3-folds \cite{eguchi} up to SL(2,Z) transformation. Base of (V) model 
is also the same as those of them.}.  
By identification of  a Mori-vector with one from elliptic fiber  i.e.,
and by taking the limit of large elliptic fiber in GKZ equations  
as $z_f \rightarrow 0 $,  
we can get a local CY3-fold \cite{chiang}. 
We would like to  use this  data  
to compactify type II/M  and to  get  5-dim. gauge theory
on  $M_4 \times S^1$ \cite{eguchi}\footnote{Work is in progress.}.

\text{\bf Acknowledgment}
\newline
I   would like to  thank  K. Mohri   for  fruitful   discussions.
I   would like to thank  S. Hosono  for    his help  in   
 calculating   Mori vectors and topological invariants in this paper 
 including the method of identification 
of phases. I  would like to  thank to N. Sakai  for useful    
discussions. 
\appendix\text{\bf { Appendix 1~ 
Linear relations among the vertices }}
\par
We list some linear relations  of one-cones 
 in the dual polyhedra,  $ Q_{(i)} $.   
They are  called U(1) charges and  have SL(5,Z) option. (
We use   charges in the left hand side.)
\newline
$\bullet$ transposed U(1) in (III)\footnote{ 
In  criterion I cases of these models,   
we can choose   U(1) charges so that    divisors  
 with the  same topological invariants in  
two models have the same transformation matrices from the common 
basis of K\"ahler cones.  Namely, for  
fixed Mori-vectors,
we   choose  U(1) charges of two models so that   
 at least $h^{1,1}$ numbers of divisors 
(i.e.,linear combinations of   basis of K\"ahler cones )  
with  same topological invariants have  common values in two models.
For example, in phase c and B by using $U(1)^{X_1}$, 
we can represent  the corresponding divisor  
in the same value of the transformation matrices, 
(see appendix 2).     
We can use this identification of divisors to 
  seek   a corresponding divisor from one model to another model 
 and to derive  matrices in criterion 2.   
 The  mappings of U(1) charges to identify 
two models are not unique and we must classify phases  
according to the correspondence of divisor representation  of the
same   topological invariants in two models  
 and decide which  mapping  we should use to compare  by case by case.}
\footnote{
$
{\small
X_1:( Q_i)^{\text {(III)}} \rightarrow   
( Q_i)^{\text {(III)}}_{X_1}
=
\begin{pmatrix}
 2Q_{2i}^{\text {(III)}}+Q_{1i}^{\text {(III)}}& \\    
 Q_{2i}^{\text {(III)}}+Q_{1i}^{\text {(III)}}& \\
 Q_{3i}^{\text {(III)}}& \\
 Q_{4i}^{\text {(III)}}+Q_{2i}^{\text {(III)}}& \\ 
 Q_{5i}^{\text {(III)}}& 
\end{pmatrix}
\quad {\text {and}}\quad
( Q_j)^{\text {(IV)} }
=
\begin{pmatrix}
 Q_{1j}& \\    
 Q_{2j}& \\
 Q_{3j}& \\
 Q_{4j}& \\ 
 Q_{5j}& 
\end{pmatrix}^{\text {(IV)}},
}
$
}

  : 
{\small
\begin{alignat*}{2}
&( Q_1,~ Q_2, ~ Q_3, 
 Q_4, Q_5, Q_6,
 Q_7,  Q_8,  Q_9 ) 
& &(  Q_1,~  Q_2, ~ Q_3, 
 Q_4,  Q_5,  Q_6,
 Q_7,  Q_8,  Q_9 )^{X_1}
\\
&(\ph 4, -1, \ph 1, \ph 6, \ph 0, \ph 2, \ph 0, \ph 0, \ph 0 ) 
& \rightarrow 
&(  \ph 12,  \ph 1, \ph 1, \,18, \ph 2, \ph 2, \ph 0, \ph 0, \ph 0 )\\
&(\ph 4,  \ph 1, \ph 0, \ph 6, \ph 1, \ph 0, \ph 0, \ph 0, \ph 0 )
&       \rightarrow
&(\ph 8, \ph 0, \ph 1, \ph  12, \ph 1, \ph 2, \ph 0, \ph 0, \ph 0 )\\
&(\ph 4,  \ph 0, \ph 0, \ph 6, \ph 0, \ph 1, \ph 1, \ph 0, \ph 0 )
&{}        \rightarrow
&(\ph 4, \ph 0, \ph 0,  \ph 6, \ph 0, \ph 1, \ph 1, \ph 0, \ph 0 )\\
&(\ph 2, -1, \ph 0, \ph 3, \ph 0, \ph 1, \ph 0, \ph 1, \ph 0 )
&{}     \rightarrow
&(\ph 6, \ph 0, \ph 0,  \ph 9, \ph 1, \ph 1, \ph 0, \ph 1, \ph 0 )\\
&(\ph 2,  \ph 0, \ph 0, \ph 3, \ph 0, \ph 0, \ph 0, \ph 0, \ph 1 ) 
&{} \rightarrow
&( \ph 2, \ph 0, \ph 0,  \ph 3, \ph 0, \ph 0, \ph 0, \ph 0, \ph 1 ).
\\
\end{alignat*}
}

{\small
\begin{alignat*}{2}
&\bullet {\text {transposed U(1) in (IV)}} 
& 
&\bullet {\text {transposed  U(1) in (V)}} :  
\\
&(  Q_1,  Q_2,  Q_3,  Q_4,  Q_5,  Q_6,
 Q_7,  Q_8,  Q_9 )
&\quad &( Q_1,  Q_2,  Q_3,  Q_4,  Q_5,  Q_6,
  Q_7,  Q_8,  Q_9 ) 
  \\
&(18, ~ 12, \po ~ 3,\po ~2, \po ~1, \po ~0, \po~ 0,\po~ 0, \po ~0 )
&\quad&(\ph 1,\ph 0, \ph 0, \ph 1, \ph 0, \ph 0, \ph 0, \ph  4,  \ph 6 )
\\
&(12,  \po ~8, \po ~2, \po ~1, \po ~0, \po ~1, \po ~0, \po ~0, \po ~0 )
&\quad&(\ph 0,  \ph 0,   \ph 1,  \ph 0,  \ph 0,   \ph 1,    \ph 0,  \ph 4,  \ph 6 )
\\
&(\po 6,  \po ~4, \po ~1, \po ~0, \po ~0, \po ~0, \po ~1, \po ~0, \po ~0 )   
&\quad&( \ph 0, \ph 1,   \ph 0, \ph 0,   \ph 1,  \ph 0,   \ph 0, \ph 4, \ph 6 )
\\
&(\po 9,  \po ~6, \po ~1, \po ~1, \po ~0, \po ~0, \po ~0, \po ~1, \po ~0 )
&\quad&(\ph 1,  \ph 0, \ph 1,  \ph 0,  \ph 1,   \ph 0,  \ph 0,  \ph 6,  \ph 9 )
\\
&(\po 3,  \po ~2,\po  ~0, \po ~0, \po ~0, \po ~0, \po ~0, \po ~0, \po ~1 ).
&\quad&(\ph 0, \ph 0, \ph 0,  \ph 0,  \ph 0,   \ph 0,    \ph 1,  \ph 2,  \ph 3  )
\end{alignat*}
}
\appendix\text{\bf { Appendix 2~ 
 Some pairs of the equivalent phases between   (III) and (IV)   }}  
There are three pairs between (III) and (IV) 
:
\newline
$\bullet$  phase c= phase B:
{\small
\begin{alignat*}{3}
&\bullet \{ J_1, \cdots,  J_5 \}
^{\text {(III)}}_{\text{c}} =
& \quad
&\bullet\{ J_1, \cdots,   J_5 \}
^{\text {(IV)}}_{\text{(B)}}=
\\
& \{ D_1, \cdots,  D_5 \}^{\text {(III)}} 
\begin{pmatrix}
 2& 1& 0  &  1&  0\\    
 2& 3& 1  &  1&  2\\
 2& 2& 0  &  1&  1\\
 1& 0& 0  &  0&  0\\ 
 1& 1& 0  &  0&  0
\end{pmatrix}^{\text {(III)}}_{\text{(c)}}
&\Rightarrow
&\{ D_1, \cdots,  D_5 \}^{\text {(IV)}}  
\begin{pmatrix}
 6& 7& 2  &  3&  4\\    
 4& 4& 1  &  2&  2\\
 2& 2& 0  &  1&  1\\
 3& 3& 1  &  1&  2\\ 
 1& 1& 0  &  0&  0
\end{pmatrix}^{\text {(IV)}}_{\text{(B)}},
\\
&\{c_2 \cdot   J_1, \cdots, c_2 \cdot  J_5   \}
^{\text {(III)}}_{\text{(c)}} 
=\{72,82,24,36,48\}
& \quad \Rightarrow 
&\{c_2 \cdot   J_1, \cdots, c_2   J_5 \}
^{\text {(IV)}}_{\text{(B)}}=
\{ 72,82,24,36,48 \},
\end{alignat*}
}
$
{\cal K}^0_{\text{ B}} ={\cal K}^0_{\text{ (c)}}= 
     6t_1^3 + 7t_1^2t_2 + 7t_1t_2^2 + 7t_2^3 + 2t_1^2t_3 + 
     2t_1t_2t_3 + 2t_2^2t_3 + 3t_1^2t_4 + 3t_1t_2t_4 + 
     3t_2^2t_4 + t_1t_3t_4 + t_2t_3t_4 + t_1t_4^2 + 
     t_2t_4^2 + 4t_1^2t_5 + 4t_1t_2t_5 + 4t_2^2t_5 + 
     t_1t_3t_5 + t_2t_3t_5 + 2t_1t_4t_5 + 2t_2t_4t_5 + 
     2t_1t_5^2 + 2t_2t_5^2.$
\newline
$\bullet$ phase d= phase C:

$
{\cal K}^0_{\text {(C)}} ={\cal K}^0_{\text {(d )}} =
     6t_1^3 + 7t_1^2t_2 + 7t_1t_2^2 + 7t_2^3 + 8t_1^2t_3 + 
     8t_1t_2t_3 + 8t_2^2t_3 + 8t_1t_3^2 + 8t_2t_3^2 + 
     8t_3^3 + 9t_1^2t_4 + 9t_1t_2t_4 + 9t_2^2t_4 + 
     9t_1t_3t_4 + 9t_2t_3t_4 + 9t_3^2t_4 + 9t_1t_4^2 + 
     9t_2t_4^2 + 9t_3t_4^2 + 9t_4^3 + 3t_1^2t_5 + 
     3t_1t_2t_5 + 3t_2^2t_5 + 3t_1t_3t_5 + 3t_2t_3t_5 + 
     3t_3^2t_5 + 3t_1t_4t_5 + 3t_2t_4t_5 + 3t_3t_4t_5 + 
     3t_4^2t_5 + t_1t_5^2 + t_2t_5^2 + t_3t_5^2 + t_4t_5^2.
$
\footnote{We can see a  mapping from (III) to (IV) from these three pairs.  
In some special cases, we might  represent the corresponding divisor 
 in the  same value of the transformation matrices from the 
basis in criterion 1, i.e.,
the mapping of  $X_1$ of U(1) charge    corresponds to this mapping 
between two models:
{\small
\begin{alignat*}{2}
X_1:
\begin{pmatrix}
 A_{1i}& \\    
 A_{2i}& \\
 A_{3i}& \\
 A_{4i}& \\ 
 A_{5i}& 
\end{pmatrix}^{\text{ (III)}}
\Rightarrow
\begin{pmatrix}
 A_{2i}& \\    
 A_{2i}& \\
 A_{3i}& \\
 A_{4i}& \\ 
 A_{5i}& 
\end{pmatrix}^{\text {(IV)}},
\begin{pmatrix}
 A_{1i}& \\    
 A_{2i}& \\
 A_{3i}& \\
 A_{4i}& \\ 
 A_{5i}& 
\end{pmatrix}^{\text {(IV)}}
=
\begin{pmatrix}
 2A_{2i}^{\text {(III)}}+A_{1i}^{\text {(III)}}& \\    
 A_{2i}^{\text {(III)}}+A_{1i}^{\text {(III)}}& \\
 A_{3i}^{\text {(III)}}& \\
 A_{4i}^{\text {(III)}}+A_{2i}^{\text {(III)}}& \\ 
 A_{5i}^{\text {(III)}}& 
\end{pmatrix}.
\end{alignat*}
}
}

$\bullet$ phase b= phase D:

$
{\cal K}^0_{\text {(D)}} ={\cal K}^0_{\text{( b)}} =
     6t_1^3 + 8t_1^2t_2 + 8t_1t_2^2 + 8t_2^3 + 2t_1^2t_3 + 
     2t_1t_2t_3 + 2t_2^2t_3 + 7t_1^2t_4 + 8t_1t_2t_4 + 
     8t_2^2t_4 + 2t_1t_3t_4 + 2t_2t_3t_4 + 7t_1t_4^2 + 
     8t_2t_4^2 + 2t_3t_4^2 + 7t_4^3 + 4t_1^2t_5 + 
     4t_1t_2t_5 + 4t_2^2t_5 + t_1t_3t_5 + t_2t_3t_5 + 
     4t_1t_4t_5 + 4t_2t_4t_5 + t_3t_4t_5 + 4t_4^2t_5 + 
     2t_1t_5^2 + 2t_2t_5^2 + 2t_4t_5^2.
$

There is another pair.  
$\bullet$ phase a = phase E
\footnote{ This pair satifies another mapping.}
{\small
\begin{alignat*}{3}
&\bullet \{ J_1, \cdots,  J_5 \}^{\text {(III)}}_{\text{(a)}} =
& \quad
&\bullet \{  J_1, \cdots,   J_5 \}^{\text {(IV)}}_{\text{(E)}}=
\\
&\{ D_1, \cdots,   D_5 \}^{\text {(III)}} 
\begin{pmatrix}
 2& 3& 2  &  4&  1\\    
 2& 2& 0  &  2&  1\\
 2& 2& 1  &  3&  1\\
 1& 1& 1  &  1&  0\\ 
 1& 1& 0  &  1&  0
\end{pmatrix}^{\text{ (III)}}_{\text{(a)}}
&\Rightarrow
&\{  D_1, \cdots,  D_5 \}^{\text {(IV)}}  
\begin{pmatrix}
 6& 7& 2  &  8&  3\\    
 4& 3& 1  &  5&  2\\
 2& 2& 0  &  2&  1\\
 3& 3& 1  &  3&  1\\ 
 1& 1& 0  &  1&  0
\end{pmatrix}^{\text {(IV)}}_{\text{(E)}},
\\
&\{c_2 \cdot  J_1, \cdots,c_2 \cdot   J_5  \}
^{\text {(III)}}_{\text{(a)}} 
=\{72,82,24,92,36\}
& \quad \Rightarrow 
&\{c_2 \cdot   J_1, \cdots, c_2 \cdot   J_5 \}^
{\text {(IV)}}_{(E)}=
\{ 72,82,24,92,36\},
\end{alignat*}
}
$
{\cal K}^0_{\text {(a)}} ={\cal K}^0_{{\text {(E)}}}= 
     6t_1^3 + 7t_1^2t_2 + 7t_1t_2^2 + 7t_2^3 + 2t_1^2t_3 + 
     2t_1t_2t_3 + 2t_2^2t_3 + 8t_1^2t_4 + 8t_1t_2t_4 + 
     8t_2^2t_4 + 2t_1t_3t_4 + 2t_2t_3t_4 + 8t_1t_4^2 + 
     8t_2t_4^2 + 2t_3t_4^2 + 8t_4^3 + 3t_1^2t_5 + 
     3t_1t_2t_5 + 3t_2^2t_5 + t_1t_3t_5 + t_2t_3t_5 + 
     3t_1t_4t_5 + 3t_2t_4t_5 + t_3t_4t_5 + 3t_4^2t_5 + 
     t_1t_5^2 + t_2t_5^2 + t_4t_5^2.
$
\appendix{\text{\bf Appendix 3~An example of equivalent phases }}
\par

Mori-vectors in phase A of (IV)  and g of (III) are given by

{\small 
\begin{align*}
\phantom{(\ph 0, \ph 0, \ph 0}\{\ell_i\}_{(\text{A})}
&\quad
\phantom{(\ph 0, \ph 0, \ph 0}\{\ell_i\}_{(\text{g})}
\\
( \ph3, \ph 2, \ph 0, \ph 0, \ph 0, \ph 0, \ph 0, \ph 0, \ph 1)
& \quad
( \ph 2, \ph 0, \ph 0, \ph 3, \ph 0, \ph 0, \ph 0, \ph 0, \ph 1)
\\
 ( \ph 0, \ph 0, \ph 0, \ph 0, \ph 1, -2, \ph 1, \ph 0, \ph 0)
& \quad
 ( \ph 0, \ph 0, \ph 1, \ph 0, \ph 1, \ph 0, -2, \ph 0, \ph 0)
\\
 (\ph 0, \ph 0, \ph 0, \ph 1, \ph 0, \ph 1, -2,\ph  0, \ph 0)
& \quad
 (\ph 0, \ph 1, \ph 1, \ph 0, \ph 0, \ph 0, \ph 0,-2, \ph 0)
\\
 ( \ph 0, \ph 0, \ph 1, \ph 0, \ph 1, \ph 0, \ph 0, -2, \ph 0)
& \quad
( \ph 0, -1, \ph 0, \ph 0, \ph 0, \ph 1, \ph 0, \ph 1, -1)
\
\\
 (\ph 0, \ph 0, \ph 0, \ph 0, -1, \ph 1, \ph 0, \ph 1, -1)
& \quad 
(\ph 0, \ph 0, -1, \ph 0, \ph 0, \ph 1, \ph 1, \ph 1, -1)
\
\end{align*}   
}

They satisfy $Q^{(A)}_i=A^{(A)}_{ij}  \ell^{(A)}_j$ etc.  
\par
We show  a   transformation matrix from phase A to phase g, 
 $M_{\text {Ag}}$ :phase A of  (IV)
$\rightarrow$ phase g of (III)  
\footnote{ We can derive the same 
matrix as  $M_{\text{(Ag)}}=M_{\text{(cg)}}^{\text{(III)}} 
M_{\text{(AB)}}^{\text{(IV)}}$  via phase B=phase c. 
}.

$
{\small
M_{\text{Ag}}^{\text{T}} := 
A_{(g)}^{\text{(IV),T}}A_{(A)}^{\text{(IV),T}}{}^{(-1)} 
=\begin{pmatrix}
 1& \ph 0&  \ph 0&  \ph 0&  \ph 0\\    
 0& \ph 0&  \ph 1&  \ph 0&  \ph 0\\
 0& \ph 0&  \ph 0&  \ph 1&  \ph 0\\
 0& -1  &  \ph 0&  \ph 0&  \ph 1\\ 
 0& \ph 1&  \ph 0&  \ph 0&  \ph 0
\end{pmatrix}
}
$


{\small
\begin{alignat*}{3}
&\bullet \{ J_1, \cdots,  J_5 \}^{\text {(III)}}_{\text{(g)}} =
& \quad
&
\bullet \{  J_1, \cdots,   J_5 \}^{\text {(IV)}}_{\text{(g)}}=
\\
&\{ D_1, \cdots,   D_5 \}^{\text {(IV)}} 
A ^{\text {(III)}}_{\text{(g)}}=
&\stackrel{X_1}\Rightarrow 
&
\{  D_1, \cdots,  D_5 \}^{\text {(IV)}}
A ^{\text {(IV)}}_{\text{(g)}}=
\\
&\{ D_1, \cdots,   D_5 \}^{\text {(IV)}} 
\begin{pmatrix}
2& 0& 1& 2  &  0&  \\    
2& 1& 1& 0  &  2&  \\
2& 0& 1& 1  &  1&  \\
1& 0& 0& 1  &  0&  \\ 
1& 0& 0& 0  &  0&  
\end{pmatrix}^{\text{ (III)}}_{\text{(g)}}
& \stackrel{X_1}\Rightarrow
&  \{  D_1, \cdots,  D_5 \}^{\text {(IV)}}
\begin{pmatrix}
6& 2& 3& 2  &  4\\    
4& 1& 2& 2  &  2  \\
2& 0& 1& 1  &  1  \\
3& 1& 1& 1  &  2  \\ 
1& 0& 0& 0  &  0  
\end{pmatrix}^{\text {(IV)}}_{\text{(g)}},
\\
&\{c_2 \cdot  J_1, \cdots,c_2 \cdot   J_5  \}
^{\text {(III)}}_{\text{(g)}} 
=\{72,24,36,24,48\}
&
 \quad \Rightarrow 
&
\{c_2 \cdot   J_1, \cdots, c_2 \cdot   J_5 \}^
{\text {(IV)}}_{(g)}=
\{72, 24,36,24,48\},
\end{alignat*}
}

We use the following identification of divisors.   
$\{ c_2 \cdot  J_i \} \ni \{72',48\} \rightarrow \{72'-48=24,48\}$, 
which is transformed to a divisor with $c_2 \cdot  J_i=24$ in 
phase g of  (III) model\footnote{ In this paper, we omitted 
most data of (V) model and mappings since the procedure is the same,
 though it is complicated due to the redundancy of K3 fiberations.}.  

{\small
\begin{alignat*}{2}
&  J_i^{\text {(IV)}} \mid_{ c_2 \cdot  J_i=72'} 
-   J_i^{\text {(IV)}}\mid_{ c_2 \cdot  J_i=48}
=
  J_i^{\text {(IV)}} \mid_{ c_2 \cdot  J_i=24}
\rightarrow 
  J_i^{\text {(III)}} \mid_{ c_2 \cdot  J_i=24}
&
\\
&=\{ D_1, \cdots,  D_5 \}^{\text {(IV)}} 
\begin{pmatrix}
\begin{pmatrix}
 6& \\    
 4& \\
 2& \\
 3& \\ 
 0& 
\end{pmatrix}
-\begin{pmatrix}
 4& \\    
 2& \\
 1& \\
 2& \\ 
 0& 
\end{pmatrix}
\end{pmatrix}^{\text {(IV)}}
=\{  D_1, \cdots,  D_5 \}^{\text {(IV)}}
\begin{pmatrix}
 2& \\    
 2& \\
 1& \\
 1& \\ 
 0& 
\end{pmatrix}^{\text {(IV)}}
&\stackrel{X_1^{(-1)}}\rightarrow
\begin{pmatrix}
 2& \\    
 0& \\
 1& \\
 1& \\ 
 0& 
\end{pmatrix}^{\text{(III)}}
\end{alignat*}
}

{\small
\begin{alignat*}{3}
&\bullet \{ J_1, \cdots,  J_5 \}^{\text {(IV)}}_{\text{(A)}} =
& \quad
&\bullet \{ J_1, \cdots,  J_5 \}^{\text {(IV)}}_{\text{(g)}} =
\\
&\{ D_1, \cdots,   D_5 \}^{\text {(IV)}} 
\begin{pmatrix}
 6& 4& 2  &  3&  6\\    
 4& 2& 1  &  2&  4\\
 2& 1& 0  &  1&  2\\
 3& 2& 1  &  1&  3\\ 
 1& 0& 0  &  0&  0
\end{pmatrix}^{\text{ (IV)}}_{\text{(A)}}
&
\stackrel{{M_{Ag}^T}J_{(A)}^T=J_{(g)}^T}\Rightarrow
&
\{  D_1, \cdots,  D_5 \}^{\text {(IV)}}  
\begin{pmatrix}
6& 2& 4& 2  &  3\\    
4& 1& 2& 2  &  2\\
2& 0& 1& 1  &  1\\
3& 1& 2& 1  &  1\\ 
1& 0& 0& 0  &  0
\end{pmatrix}^{\text {(IV)}}_{\text{(g)}},
\\
&\{c_2 \cdot  J_1, \cdots,c_2 \cdot   J_5  \}
^{\text {(IV)}}_{\text{(A)}} 
=\{72,48,24,36,72'\}
&
 \quad \Rightarrow 
&
\{c_2 \cdot   J_1, \cdots, c_2 \cdot   J_5 \}^
{\text {(IV)}}_{(g)}=
\{ 72,24,48,24,36\},
\end{alignat*}
}

This  matrix  is   a one to one mapping 
of  topological invariants.    
These phases are   
topologically equivalent.
The Gromov-Witten invariants transform 
{\small
\newline
\[
\begin{array}{rrrrrrrr}
N(1, 0, 0, 0, 1)_{(\text{A})}&=&252 & \rightarrow & 
N(1, 0, 0, 1, 0)_{\text{(g)}}&=&252,
\\
N(1, 0, 0, 1, 1)_{(\text{A})}&=&252& \rightarrow & 
N(1, 0, 1, 1, 0)_{\text{(g)}}&=&252, 
\\
N(1, 1, 0, 1, 1)_{(\text{A})}&=&252& \rightarrow & 
N(1, 0, 1, 0, 1)_{\text{(g)}}& =& 252.
\\
N(1, 1, 0, 0, 1)_{(\text{A})}&=&252& \rightarrow & 
N(1, 0, 0, 0, 1)_{\text{(g)}}& =& 252.
\\
N(1, 1, 1, 1, 1)_{(\text{A})}&=&252& \rightarrow & 
N(1, 1, 1, 0, 1)_{\text{(g)}}& =& 252.
\\
N(1, 1, 1, 0, 1)_{(\text{A})}&=&252& \rightarrow & 
N(1, 1, 0, 0, 1)_{\text{(g)}}& =&252.
\nn
\end{array}
\]
}
\appendix{\text{\bf Appendix~4~ 
The other examples of  criterion 2}}
\par 
The list of some transformation matrices of the 
topological invariants of the equivalent phases.
The   mappings   of $c_2 \cdot  J_i$ in 
the equivalent phases :
\newline
$\bullet$ $ g \rightarrow \alpha_{10},  B \rightarrow$  e,  
 D $\rightarrow$ f  :   $\{ 24,48 \} \rightarrow \{ 24,24=48-24 \}$
\newline
$\bullet$  
$\alpha_3 \rightarrow $ f:   
$\{ 72,82,82 \} \rightarrow \{ 72,82,92=82+82-72 \}$
\newline
$\bullet$  $\alpha_{18}  \rightarrow$ C  :  
$\{ 72,82,92,92' \} \rightarrow \{ 72,82,92,102=92+92-82' \}$ .
\newline
$\bullet$ h and E  :  
$ \{ 92,82, 82' \} \rightarrow  \{ 92,82,72=82'-92+82 \} $
{\small
\begin{alignat*}{2}
 M_{\text {(ce)}}^{\text T}&=
\begin{pmatrix}
 1 & \ph 0 & \ph 0 & \ph 0 & \ph 0 \\ 
 0 & \ph 0 & \ph 0    & \ph 1 & \ph 0 \\ 
 0 & \ph 0 & \ph 1 & \ph 0 & \ph 0 \\
 0 & \ph 0 & -1  & \ph 0 & \ph 1 \\ 
 0 & \ph 1   & \ph 0 & \ph 0   & \ph 0
\end{pmatrix},
& \quad
 M_{\text {(bf)}}^{\text T}&=
\begin{pmatrix}
 1 &  \ph 0 & \ph 0 & \ph 0 & \ph 0\\
 0 &  \ph 0 & \ph 0 & \ph 1 & \ph 0\\ 
 0 & \ph 0 & \ph 1 & \ph 0 & \ph 0\\ 
 0 & \ph 0 & \ -1 & \ph 0 & \ph 1\\ 
 0 & \ph 1 & \ph 0 & \ph 0 & \ph 0
\end{pmatrix},
\end{alignat*}
}
{\small
\begin{alignat*}{2}
 M_{(\alpha_3 {\text {f}})}^{\text {T}}&= 
\begin{pmatrix}
 0 & \ph 0 & \ph 1 & \ph 0 & \ph 0\\ 
 0 & \ph 1 & \ph 0 & \ph 0 & \ph 0\\ 
 0 & \ph 0 & \ph 0 & \ph 1 & \ph 0\\ 
 0 & \ph 0 & \ph 0 & \ph 0 & \ph 1\\ 
 1 & \ph 1 & \ -1 & \ph 0 & \ph 0
\end{pmatrix},
&\quad
 M_{(\alpha_{18}{\text {d}})}^{\text T} &= 
\begin{pmatrix}
 0 & \ph 0 & \ph  0 & \ph  1 &\ph  0\\ 
 0 & \ph 0 & \ph 1 &  \ph 0 & \ph 0\\ 
 1 & \ph 0 & \ph 0 &  \ph 0 & \ph 0\\
 1 & \ph 1 & \ -1 & \ph 0 & \ph 0\\ 
 0 & \ph 0 & \ph  0 & \ph  0 & \ph 1
\end{pmatrix}
\\
 M_{({\text {g}}\alpha_{10})}^{\text T} &= 
\begin{pmatrix}
 1 & \ph 0 & \ph 0 & \ph 0 &  \ph 0\\ 
 0 & \ph 0 & \ph 1 &  \ph 0 & \ph 0\\ 
 0 & \ph 0 & \ph 0 & \ph 1 & \ph 0\\
 0 & \ -1 & \ph 0 & \ph 0 & \ph 1\\ 
 0 & \ph 1  & \ph 0 & \ph 0 & \ph 0
\end{pmatrix},
&\quad
 M_{{\text {( ha )}}}^{\text {T}} &= 
\begin{pmatrix}
 0 & \ph 0 & \ph 1 & \ -1 &  \ph 1\\ 
 0 & \ph 0 & \ph 1 &  \ph 0 & \ph 0\\ 
 0 & \ph 1 & \ph 0 & \ph 0 & \ph 0\\
 0 & \ph 0 & \ph 0 & \ph 1 & \ph 0\\ 
 1 & \ph 0  & \ph 0 & \ph 0 & \ph 0
\end{pmatrix}.
\end{alignat*}
}
\newline
For the others, they are not equivalent
\footnote
{
$\bullet$ $ \{  A,g, \alpha_{10} \} \not= \{ B, C, D, E\}$.   
$\{ c_2 \cdot  J_i\}_{A, g, \alpha_{10}}$   
are only multiple of 12.  
$\{ c_2 \cdot  J_i \}_{ B, C, D, E}$ are not.
\newline
\indent
$\bullet$  $B   \not= E $,   
$ \{ c_2 \cdot  J\}_B \ni \{ 48 \}$,
$   \{ c_2 \cdot  J \}_E \not\ni \{ 48 \}   $. 
To make 48  from 24,  $72'$  is necessary.   
\newline
\indent
$\bullet$  $B,D,E    \not= C $,  
$ \{ c_2 \cdot  J\}_{B,D,E} \ni \{ 24 \}$,
$   \{ c_2 \cdot  J \}_C  \not\ni \{ 24 \} $
\newline
\indent
$\bullet$   $B, E    \not= D $,  
$ \{ c_2 \cdot  J\}_{B,E} \ni \{36\}$,
$ \{ c_2 \cdot  J \}_D \not \ni \{ 36 \}$.   }. 
\appendix\text{\bf Appendix~5 Topological data of ${\cal K}^0$ and  
$c_2 \cdot  J$  }
\newline 
We list the topological  data of three models
\footnote{We also list the ring data in table 8. 
Topological invariants are calculated by the method in \cite{hosono2}.
 The author thanks S. Hosono for his help in the calculation.}. 

(III) There are 8 phases.
We list the data of four  phases
\begin{itemize}
\item
 phase e:  
$
{\cal K}^0_{({\text e})} = 6t_1^3 + 3t_1^2t_2 + t_1t_2^2 + 2t_1^2t_3 + 
     t_1t_2t_3 + 2t_1^2t_4 + t_1t_2t_4 + t_1t_3t_4 + 
     7t_1^2t_5 + 3t_1t_2t_5 + t_2^2t_5 + 2t_1t_3t_5 + 
     t_2t_3t_5 + 2t_1t_4t_5 + t_2t_4t_5 + t_3t_4t_5 + 
     7t_1t_5^2 + 3t_2t_5^2 + 2t_3t_5^2 + 2t_4t_5^2 + 7t_5^3;$
\newline
$c_2 \cdot  J_{\text{(e)}} = \{72, 36, 24, 24, 82\}$;
\item
 phase f:  
$
{\cal K}^0_{({\text f})} = 
     6t_1^3 + 7t_1^2t_2 + 7t_1t_2^2 + 7t_2^3 + 2t_1^2t_3 + 
     2t_1t_2t_3 + 2t_2^2t_3 + 2t_1^2t_4 + 2t_1t_2t_4 + 
     2t_2^2t_4 + t_1t_3t_4 + t_2t_3t_4 + 8t_1^2t_5 + 
     8t_1t_2t_5 + 8t_2^2t_5 + 2t_1t_3t_5 + 2t_2t_3t_5 + 
     2t_1t_4t_5 + 2t_2t_4t_5 + t_3t_4t_5 + 8t_1t_5^2 + 
     8t_2t_5^2 + 2t_3t_5^2 + 2t_4t_5^2 + 8t_5^3;$
\newline
$c_2 \cdot  J_{\text{(f)}}$ = \{72, 82, 24, 24, 92\};
\item
 phase g:  
$
{\cal K}^0_{({\text g})} = 
     6t_1^3 + 2t_1^2t_2 + 3t_1^2t_3 + t_1t_2t_3 + 
     t_1t_3^2 + 2t_1^2t_4 + t_1t_2t_4 + t_1t_3t_4 + 
     4t_1^2t_5 + t_1t_2t_5 + 2t_1t_3t_5 + 2t_1t_4t_5 + 
     2t_1t_5^2;$
\newline
$c_2 \cdot  J_{\text{(g)}}$ = \{72, 24, 36, 24, 48\}; 
\item
 phase h:  
$
{\cal K}^0_{({\text h})} = t_1^2t_3 + t_1t_2t_3 + 3t_1t_3^2 + 2t_2t_3^2 + 
     7t_3^3 + t_1^2t_4 + t_1t_2t_4 + 3t_1t_3t_4 + 
     2t_2t_3t_4 + 8t_3^2t_4 + 3t_1t_4^2 + 2t_2t_4^2 + 
     8t_3t_4^2 + 8t_4^3 + t_1^2t_5 + t_1t_2t_5 + 
     3t_1t_3t_5 + 2t_2t_3t_5 + 8t_3^2t_5 + 3t_1t_4t_5 + 
     2t_2t_4t_5 + 8t_3t_4t_5 + 8t_4^2t_5 + 3t_1t_5^2 + 
     2t_2t_5^2 + 8t_3t_5^2 + 8t_4t_5^2 + 7t_5^3;$
\newline
$c_2 \cdot  J_{\text{(h)}}$ = \{36, 24, 82, 92, 82\};
\end{itemize}
(IV) There are five phases  called  A, B, C, D, E.
We  list only the phase A since the other four phase data coincide with
the data of (III).
\newline
\begin{itemize}
\item
phase A:
$
{\cal K}^0_{({\text A})} = 6t_1^3 + 4t_1^2t_2 + 2t_1t_2^2 + 2t_1^2t_3 + 
     t_1t_2t_3 + 3t_1^2t_4 + 2t_1t_2t_4 + t_1t_3t_4 + 
     t_1t_4^2 + 6t_1^2t_5 + 4t_1t_2t_5 + 2t_1t_3t_5 + 
     3t_1t_4t_5 + 6t_1t_5^2;$
\newline
$c_2 \cdot  J_{\text{(A)}}$   = \{72, 48, 24, 36, 72'\};
\end{itemize}


(V) There are 18 phases.
We list only the data of phases 3, 10  and 18, since the  other 15 data 
coincide with those in case (III).
\begin{itemize}
\item
 phase $\alpha_3$:  
$
{\cal K}^0_{(\alpha_3)}= 
     7t_1^3 + 8t_1^2t_2 + 8t_1t_2^2 + 7t_2^3 + 7t_1^2t_3 + 
     8t_1t_2t_3 + 7t_2^2t_3 + 7t_1t_3^2 + 7t_2t_3^2 + 
     6t_3^3 + 2t_1^2t_4 + 2t_1t_2t_4 + 2t_2^2t_4 + 
     2t_1t_3t_4 + 2t_2t_3t_4 + 2t_3^2t_4 + 2t_1^2t_5 + 
     2t_1t_2t_5 + 2t_2^2t_5 + 2t_1t_3t_5 + 2t_2t_3t_5 + 
     2t_3^2t_5 + t_1t_4t_5 + t_2t_4t_5 + t_3t_4t_5;$
\newline
$c_2 \cdot  J_{(\alpha_3)}$ = \{82, 82, 72, 24, 24\};

\item
  phase $\alpha_{10}$: 
$
{\cal K}^0_{(\alpha_{10})}= 6t_1^3 + 3t_1^2t_2 + t_1t_2^2 + 2t_1^2t_3 + 
     t_1t_2t_3 + 2t_1^2t_4 + t_1t_2t_4 + t_1t_3t_4 + 
     2t_1^2t_5 + t_1t_2t_5 + t_1t_3t_5 + t_1t_4t_5;$
\newline
$c_2 \cdot  J_{(\alpha_{10})}$ = \{72, 36, 24, 24, 24\};
\item
  phase $\alpha_{18}$: 
$
{\cal K}^0_{(\alpha_{18})}=
     8t_1^3 + 9t_1^2t_2 + 9t_1t_2^2 + 8t_2^3 + 8t_1^2t_3 + 
     9t_1t_2t_3 + 8t_2^2t_3 + 8t_1t_3^2 + 8t_2t_3^2 + 
     7t_3^3 + 8t_1^2t_4 + 9t_1t_2t_4 + 8t_2^2t_4 + 
     8t_1t_3t_4 + 8t_2t_3t_4 + 7t_3^2t_4 + 8t_1t_4^2 + 
     8t_2t_4^2 + 7t_3t_4^2 + 6t_4^3 + 3t_1^2t_5 + 
     3t_1t_2t_5 + 3t_2^2t_5 + 3t_1t_3t_5 + 3t_2t_3t_5 + 
     3t_3^2t_5 + 3t_1t_4t_5 + 3t_2t_4t_5 + 3t_3t_4t_5 + 
     3t_4^2t_5 + t_1t_5^2 + t_2t_5^2 + t_3t_5^2 + t_4t_5^2;$
\newline
$c_2 \cdot  J_{(\alpha_{18})}$ = \{92, 92', 82, 72, 36\};
\end{itemize}

\appendix\text{\bf Appendix 6  Gromov-Witten inv. of CY3-fold with 
(4,214) and    a tensor  }
\newline
The Mori vectors in {\bf CFPR} model side  
corresponding to   phase 6 in \cite{theisen} are  given by   

\begin{alignat*}{2}
&( ~a_1,~ a_2,  ~a_4, ~a_5, ~a_6, ~a_7, ~a_8,~ a_9 & )& \quad \\ 
&(\ph 2,    \ph 1,  \ph 3,  \ph 0,  \ph 0,  \ph 1, -1,     \ph 0 &  )&=\ell 
_1 \quad  \\
&(\ph 0,   \ph 1,  \ph 0, \ph 1, \ph 0,  \ph 0, \ph 0,   -2 & )  &
=\ell_2 \quad 
    \\
&(\ph 0,    \ph 0,  \ph 0,  \ph 0,  \ph 1,  \ph 1, \ph 0,   -2  &)&
=\ell_3   \quad 
 \\
&(\ph 0,    -1,  \ph 0,  \ph 0,  \ph 0,  -1, \ph 1,   \ph 1  &) &
=\ell_4.  \quad
\end{alignat*}
The ${\cal K}^0_{ijk}{}_{II}$ and $C_2\cdot J_i$ of phase 6  is given  in \cite{theisen}.
\begin{align}
{\cal  K}^0
& =
7t_4^3 + 2t_4^2t_2 + 2t_4^2t_3 + t_4t_2t_3 + 8t_4^2t_1
\nonumber \\ 
&+ 2t_4t_2t_1 + 2t_4t_3t_1 + t_2t_3t_1 + 
8t_4t_1^2 + 2t_2t_1^2 + 2t_3t_1^2 + 8t_1^3,\nonumber \\ 
& C_2 \cdot J=\{82,24,24,92\}. 
\end{align}
In this case, $t_2$ and $t_3$  are symmetric  
\footnote
{${\cal K}^0_{ijk}{}_{ \text { II} }$ of  phase f 
 with  $(h^{1,1},h^{2,1})=(5,185)$  
 can be truncated to the one in  6 phase case   by setting $t_1=0$ 
and replacing  $t_{i+1} \rightarrow t_i$ for $i=2 \cdots 5$, 
which is given by \cite{theisen}.}.
\[
 K^{(g=0)}_{II}{}^{\text NP}
=   
\frac{1}{(2 \pi)^3}\sum_{n_1,n_2,n_3,n_4}
N (n_1,n_2,n_3,n_4){\text {Li}}_3(\Pi^4_{i=1}q_i^
{n_i}).
\]
\par 
The  reductions of the  Mori vectors of CY3-fold to those of  
 $B_8$, $S-T-U$ model and $F_2$ 
with a blow up point are as follows: 
\newline
$\bullet$
$(\ell_1,\ell_2,\ell_3,\ell_4) \rightarrow (\ell_1+\ell_4, \ell_2-\ell_3),
(\ell_1+\ell_4, \ell_2+\ell_4),(\ell_1+\ell_4, \ell_3+\ell_4)$ 
for $B_8$, 
\newline
$\bullet$
$(\ell_1,\ell_2,\ell_3,\ell_4) \rightarrow (\ell_1+\ell_4, \ell_2,\ell_3)$
for S-T-U model with (3,243), 
\newline
$\bullet$
$(\ell_1,\ell_2,\ell_3,\ell_4)\rightarrow{}_{(T^2\rightarrow \infty)} 
(0, \ell_2^\prime,\ell_3^\prime,\ell_4^\prime)$
for  $F_2$ with a blow up point.   
\par
\par 
To take   the heterotic string side, we use the linear transformations 
 from $t_i$ to $ S,T,U,V $ given in \cite{theisen}  :
$t_1=V,~ t_2=T-U,~ t_3=S-U,~ t_4=U-V$.
After taking a weak couplig limit  
 such as $t_3 \rightarrow \infty$ and   $q_3=n_3=0$,  
the only sequences with $n_1 \geq n_2$ seem to remain. 

The  sequences with $n_1=n_2,n_3=0$  are  represented 
by  $Z_{0 ;n}^{\text inst}{}^{B_8}(\tau )$ where  
$Z_{g ;n}^{\text inst}{}^{B_8}(\tau )
=\Sigma_{k\mid n}\mu(k)k^{-3}Z_{g,{\frac{n}{k}}}{}^{B_8}(k\tau)$
and  $\mu(k)$  is a M\"obius function.   
 \newline
$Z_{0;1}^{B_8}=\frac{1}{\eta^{12}}E_4$,
$Z_{0;2}^{B_8}=\frac{1}{24\eta^{24}}E_4(2E_6+ E_2 E_4)$,
\newline
$Z_{0;3}^{B_8}=\frac{1}{15552\eta^{36}}E_4
(109E_4^3+197E_6^2+216E_2E_4E_6+54E_2^2E_4^2)$,
etc.
\newline
Therefore, they seem to be represented by  the  Dedekind eta function, 
$\eta(\tau )$ and the  Eisenstein series, $E_i $  only. 
\newline 
$Z_{0;n}^{\text inst}{}^{B_8}=Z_{0;n,m}^{\text inst}{}^{B_8}q^m$:
\newline
$Z_{0;1}^{\text inst}{}^{B_8}=1+252q+5130q^2+\cdots$
\newline  
$Z_{0;2}^{\text inst}{}^{B_8}=-9252q^2-673760q^3+\cdots$
\newline
$Z_{0;3}^{\text inst}{}^{B_8}=948628q^3+115243155q^4+\cdots$
, etc.
\newline
The examples of the 
Gromov-Witten invariants  with $n_1 > n_2$ and  $n_3=0$ that    
  are not represented by $Z^{\text inst}_{0;n}{}^{B_8}$ 
  are N(2,1,0,2)=265968, N(3,1,0,3)=162273760 and  N(3,1,0,2)=1739160 etc 
\footnote{
However, 265968 and 162273760   exist in 
the Gromov-Witten invariants of CY3-fold with (4,214) 
and   $(k_1,k_2)=(11,13)$.     
It  is    $\chi_0$  with a extra vector multiplet
and K3={\bf P}$^3(1,1,3,5)[12]$ fiber that is  given by \cite{curio}.
In $\chi_0$ case, both  
$\Sigma_{n_4} N(n_1,0,n_3,n_4)$ and 
$\Sigma_{n_4} N(n_1,n_2,0,n_4)$
lead to the   coefficient of $\frac {2 E_4 E_6}{\eta^{24}}$ expansion
though S and T are not symmetric,
because  they reduce to  $(h^{1,1},h^{1,2})=(3,243)$ case 
where S and T are symmetric.  
N(2,0,1,4)=265968 and N(3,0,1,6)=162273760    
correspond  to the limit of $T \rightarrow  \infty$.  
They are in the contribution from the non-perturbative 
vector multiplet for taking  S as the dilaton.}    
\footnote{1739160 exists in the Gromov-Witten invariants of 
$(h^{1,1},h^{1,2})=(5,185)$ with two tensors case such as 
the phase f. In this case, 
almost Gromov-Witten invariants are represented by those of 
$B_8$. The others relate to those of the phase 16 
of the list that is given by \cite{theisen}. The phase 16 has 
a 6-dim tensor and a 6-dim vector\cite{mmmabe}.}.
\footnote{The phase 14 of the list that is given by \cite{theisen} 
coinsides with the perturbative coupling such as  
$\frac { E_{4,1}(r_1,\tau) E_{6,1}(r_2,\tau) }{\eta^{24}}
+\frac { E_{4,1}(r_2,\tau ) E_{6,1}(r_1,\tau )  }{\eta^{24}}$. 
which is the N=2 model with two  6-dim. vector multiplets\cite{mmmabe}. }  
\newline
$
\lim_{n_3=0} K^{(g=0)}_{II}{}^{\text NP}
 = 
 420 (\Sigma_n Li_3((q_1q_4)^n)+\Sigma_n Li_3((q_1q_2q_4)^n))  
$
\newline
$
+\Sigma_{n,m} Z^{B_8}{}^{\text inst}_{0;n,m}
(Li_3(q_1^n (q_1q_2 q_4)^m)
+\Sigma_{n,m} Z^{B_8}{}^{\text inst}_{0;n,m}
Li_3((q_1q_2q_4^2)^n(q_1q_2q_4)^m)
$
\newline
$
+\Sigma_{n,m} Z^{B_8}{}^{\text inst}_{0;n,m}
Li_3(q_1^n (q_1 q_4)^m)
+\Sigma_{n,m} Z^{B_8}{}^{\text inst}_{0;n,m}
Li_3((q_1q_2q_4)^n (q_1 q_4)^m))
$
\newline
$
+N (2,1,0,2){\text {Li}}_3(q_1^2q_2q_4^2)
+N (3,1,0,2){\text {Li}}_3(q_1^3q_2q_4^2)
+N (3,1,0,3){\text {Li}}_3(q_1^3q_2q_4^3)+\cdots.
$

The partition function of the model 
with a nonperturbative vector  for $T \rightarrow  \infty$ will be   
 also  represented by the   quasi modular forms and the 
 character of the Kac-Moody algebra including 
$E_2$ and a Wilson line. By examining their relations and  
taking an appropriate limit, 
the partition function with a tensor will be also represented 
in the quasi modular forms and the characters.

{\small\allowdisplaybreaks
\begin{table}[h]
\[
\begin{array}{|l|l||l|l||l|l||l|l||}
\hline
           & {\text{models}}
&  (I)   &G_1=I      
&  (II)  &G_1=A_1   
&  (II)  &G_1=A_2   
\\ 
\hline
n^0  & G_2   
&  (h^{1,1},  h^{2,1}) & {\text{CY3 weight}}    
&   (h^{1,1},  h^{2,1}) & {\text{CY3 weight}} 
&  (h^{1,1},  h^{2,1})& {\text{CY3 weight}}
\\ 
\hline \hline
0        & I             
& (3,243)      &     
& (4,214)      &    
& (5,197)     & 
\\ \hline
 2    &I             
&(3,243) & (1,1,2,8,12)     
&(4,190) & (1,1,2,6,10)
&(5,161) & (1,1,2,6, 8)
\\ \hline
 3    &  A_2   
& (5,251)      &(1,1,3,10,15)     
& (6,186)      & (1,1,3,7,12)
& (7,151)      & (1,1,3,7,9)
\\ \hline
 4    &  D_4   
& (7,271)    & (1,1,4,12,18)     
& (8,194)    & (1,1,4,8,14) 
& (9,153)    & (1,1,4,8,10)
\\ \hline
 6    &  E_6   
&(9,321)  & (1,1,6,16,24)     
&(10,220) & (1,1,6,10,18) 
&(11,167) & (1,1,6,10,12)
\\ \hline
 8    &  E_7   
&(10,376)  & (1,1,10,24,36)     
&(11,267)  & (1,1,10,14,26) 
&(12,186)  & (1,1,10,14,16)
\\ \hline
 12            &  E_8   
&(11,491)      & (1,1,12,28,42)    
&(12,318)      & (1,1,12,16,30) 
&(13,229)      & (1,1,12,16,18) 
\\ \hline
\end{array}
\]
\caption{ type IIA-heterotic string duality :
Hodge and instanton numbers of CY3s in (I) and (II) in A-chain  }
\end{table}
}
{\small\allowdisplaybreaks
\begin{table}[h]
\[
\begin{array}{|l|l|l|l|l|l|l|l|l|l|}
\hline 
n^0     &G_2   & h^{1,1}& h^{2,1}& k_1& k_2  
& n_ T^0&\Delta n_T& n_T& K3~ {\text { fiber}}  
\\ 
\hline \hline
 0    & I             &3      &243     & 12 & 12   &  1&0&1     
&\text{\bf  P}^3(1,1,4,6)[12]\\ \hline
 2    &I             &3       &243     & 12+2 & 12-2&  1&0&1 
&\text{\bf  P}^3(1,1,4,6)[12]\\ \hline
 3    &  A_2   &5       &251     & 12+3 &12-3 &  1&0&1 
& \text{\bf P} ^3(1,2,6,9)[18]\\ \hline
 4    &  D_4   &7      &271     & 12+4 & 12-4 &  1&0&1
& \text{\bf P}^3(1,2,6,9)[18] \\ \hline
 6    &  E_6   &9      &321     & 12+6 & 12-6 &  1&0&1 
 & \text{\bf P}^3(1,3,8,12)[24]\\ \hline
 8    &  E_7   &11      &376     & 12+8 & 12-8 &  1&0&1  
& \text{\bf P}^3(1,4,10,15)[30]\\ \hline
 12   &  E_8   &12      &491     & 12+12 & 12-12 &  1&0&1
& \text{\bf P}^3(1,5,12,18)[36] \\ \hline
\end{array}
\]
\caption{ Hodge and instanton numbers of CY3s in (I)  }
\end{table}
}
{\small\allowdisplaybreaks
\begin{table}[h]
\[
\begin{array}{|l|l|l|l|l|l|l|l|l|l|}
\hline 
n^0     &G_2  \times G_1  & h^{1,1}& h^{1,2}& k_1& k_2  \\ 
\hline
 0  &  I \times A_1     &4      &214     & 12 & 12       \\ \hline
 1  &  I \times A_1     &4      &202     & 12+1 & 12-1  \\ \hline
 2  &  A_2 \times A_1   &4      &190     & 12+2 & 12-2  \\ \hline
 6  &  E_6  \times A_1  &6      &220     & 12+6 & 12-6          \\ \hline
 8  &  E_7  \times A_1  &11     &251     & 12+8 & 12-8          \\ \hline
 10 &  E_8  \times A_1  &14     &284     & 12+10 & 12-10          \\ \hline
\end{array}
\]
\caption{ Hodge and instanton numbers of CY3s in (II) with $ G_1=A_1$ 
in A series  }
\end{table}
}

{\small\allowdisplaybreaks
\begin{table}[h]
\[
\begin{array}{|l|l|l|l|l|l|l|l|l|l|}
\hline 
\Delta n_T     & G_2   & h^{1,1}& h^{2,1}& k_1& k_2  
& n_ T^0 & n_T &\Delta h^{1,1}&\Delta h^{2,1} \\ \hline \hline 
0    &  I             &3      &243     & 12 & 12   &  1&1&0&0   
  \\ \hline
2    & I             &5       &185     & 12 & 12-2&  1&3&2&-58 \\ \hline
3    &  A_2   &8       &164     & 12 &12-3 &  1&4 &3&-87\\ \hline
4    & D_4   &11      &155     & 12 & 12-4 &  1&5 &4&-116\\ \hline
6    & E_6   &15      &147     & 12 & 12-6 &  1&7&6&-174 \\ \hline
8    & E_7   &18      &144     & 12 & 12-8 &  1&9&8&-132 \\ \hline
12   & E_8   &23      &143     & 12 & 12-12 &  1&13&12&-348 \\ \hline
\end{array}
\]
\caption{The Hodge and instanton numbers in (III)/(IV) }
\end{table}
}
{\small\allowdisplaybreaks
\begin{table}[h]
\[
\begin{array}
{|l|l|l|l|l|l|}
\hline
\multicolumn{3}{|c|}{\text{CY3s in  (III)}} &
\multicolumn{3}{|c|}{ \text{CY3s in (IV)}}\\
\hline
\Delta n _T  &  \text{K3 }   & \text{K3 }
 & s         &  \text{weight}     & \text{K3 }
\\ \hline
0 &   \text{\bf P}^3(1,1,4,6)[12] &\text{\bf  P}^3(1,1,4,6)[12]
& 1 &  & \text{\bf P}^3(1,1,4,6)[12] 
\\ \hline
2 & \text{\bf P}^3(1,1,4,6)[12]   & \text{\bf P}^3 (1,1,4,6)[12]
&2&(1,1,2,8,12)&\text{\bf P}^3(1,1,4,6)[12] 
\\ \hline
3 &\text{\bf P}^3(1,1,4,6)[12] 
 & \text{\bf P} ^3(1,2,6,9)[18]
&3&(1,2,3,12,18)
&\text{\bf P}^3(1,1,4,6)[12]  
\\ \hline
4 & \text{\bf P}^3(1,1,4,6)[12]  & \text{\bf P}^3(1,2,6,9)[18]
&4& (1,4,5,20,30)&\text{\bf P}^3(1,1,4,6)[12]  
\\ \hline
6 & \text{\bf P}^3(1,1,4,6)[12] 
 & \text{\bf P}^3(1,3,8,12)[24]&6& (1,6,7,28,42)
& \text{\bf P}^3(1,1,4,6)[12]
 \\ \hline
8 & \text{\bf P}^3(1,1,4,6)[12]
  & \text{\bf P}^3(1,4,10,15)[30]&8& (1,8,9,36,54)&
\text{\bf P}^3(1,1,4,6)[12]  
\\ \hline
12 & \text{\bf P}^3(1,1,4,6)[12]
 & \text{\bf P}^3(1,5,12,18)[36]&12&(1,12,13,52,78) & 
\text{\bf P}^3(1,1,4,6)[12]
 \\ \hline
\end{array}
\]
\caption{  The type of K3  sub dual polyhedra contained in    (III) and case (IV)}
\end{table}
}
{\small\allowdisplaybreaks
\begin{table}[h]
\[
\begin{array}{|l|l|l|l|l|l|l|l|l|l|}
\hline 
s     &G_2 \times G_1   & h^{1,1}& h^{2,1}  
& \Delta h^{1,1}&\Delta h^{2,1}& k_1&k_2&\Delta n_T  
\\ 
\hline \hline
 3  & I \times   A_1  &9  &129  &5  & -85   &  & & \\ \hline
 2  & I \times   A_1  &6  &144  & 2 & -58=-29\times 2   &  12+1&12-1-2&2 \\ \hline
 1  & A_1 \times A_1  &4  &190  & 0 & 0               &  12+2&12-2&0 \\ \hline
 5  & E_6 \times A_1  &16 &118  & 6 & -102=-17\times 6  &  12+6-6&12-6& 6 \\ \hline
 7  & E_7 \times A_1  &19 &115  & 8 & -136=-17\times 8  &  12+8-8&12-8& 8\\ \hline
 9  & E_8 \times A_1  &24 &114  &10 & -170=-17\times 10 &  12+10-10&12-10&10 \\ \hline
\end{array}
\]
\caption{The Hodge numbers of $ P^4$(1,s,(1+s)(1,3,5))[10s] 
and the relation of heterotic duality for (VI)with $ G_1=A_1$ in A series.} 
\end{table}
}
{\small\allowdisplaybreaks
\begin{table}[h]
\[
\begin{array}{|c|l|l|}\hline 
{\rm model }     &   \# \{\text{K3 fibrations}\} 
&   \# \{\text{phases by triangulation}\}       \\ \hline \hline 
{(\I}^\dagger)      &    \{0,1,2\}   
& \text{\po 8 phases  }  \\ \hline
{(\III)}            &    \{0,1,2\}   
& \text{\po 8 phases  labeled by}\ {\text a}, 
{\dots}, {\text h} \\ \hline
{(\IV)}             &    \{0,1\}     
& \text{\po 5 phases labeled by}\  
\text{A},{\dots}, \text{E}   \\ \hline
{(\V)}              &    \{0,1,2,3\} 
& \text{18 phases labeled by}\ \alpha_1, {\dots}, \alpha_{18}\\ \hline
\end{array}
\]
\caption{ The number of K3 fibrations and the phases specified by the  
triangulations in four models}
\end{table}
}
{\small\allowdisplaybreaks
\begin{table}[h]
\[
\begin{array}{|l|l|l|l|l|l|l|l|}
\hline \hline 
{\text{ No~ of}}~B_8~ {\text {in (A)}}   
 &N= 252   &N= -9252&N=848628  \\ \hline
\hline 
1\quad \{n_i\}  &  10001   &20002      &30003       
  \\ \hline
2 \quad\{n_i\}   & 10011   &20022      &30033      \\ \hline
3 \quad \{n_i\}  & 11101   &22202      &33303      \\ \hline
4  \quad\{n_i\}  & 11111   &22222      &33333      \\ \hline
5\quad \{n_i\}   & 11001   &22002      &33003      \\ \hline
6 \quad \{n_i\}  & 11011   &22022      &33033     \\ \hline
 \hline
{\text{ No~ of}}~B_8~ {\text{ in (6)}}   
&N= 252   &N= -9252&N=848628  \\ \hline
\hline 
1\quad \{n_i\}   & 1000   &2000      &3000     \\ \hline
2\quad \{n_i\}   & 1012   &2024      &3036     \\ \hline
3\quad \{n_i\}   & 1102   &2204      &3306     \\ \hline
\end{array}
\]
\caption{
$N(\{n_i\})$\   of\  phase\ A \ with \ $(h^{1,1},h^{2,1})=(5,185)$ 
\ which denote $B_8$   \  and \ $N(\{n_i\})$\   of\  phase\ 6 \ with \ 
$(h^{1,1},h^{2,1})=(4,214)$ \ which denote $B_8$  }
\end{table}
}

{\small\allowdisplaybreaks
\begin{table}[h]
\[
\begin{array}{|l|l|l|l|l|l|l|l|}
\hline \hline 
{\text{ No~ of}} ~Z^{\text {inst}{B8} }_{0;0,m} ~ {\text{ in (6)}}   
    &N=252&N=5130  \\ \hline
\hline 
1\quad \{n_i\}           &1000      &2001       
  \\ \hline
2 \quad\{n_i\}           &1000      &2011      \\ \hline
3 \quad \{n_i\}          &1000      &2101      \\ \hline
4  \quad\{n_i\}          &1012      &2013      \\ \hline
5  \quad\{n_i\}          &1102      &2103      \\ \hline
6\quad \{n_i\}           &1102      &2203      \\ \hline
7 \quad \{n_i\}          &1012      &2023      \\ \hline
 \hline
\end{array}
\]
\caption{
$N(\{n_i\})$\   of\  phase\ 6 \ with \ $(h^{1,1},h^{2,1})=(4,214)$ 
\ which denote $B_8$   }
\end{table}
}

{\setlength{\textwidth}{10cm}
\begin{table}[h]
\[
\begin{array}{|c|c|c|c|}
\hline 
\sharp\{\text{K3 fibrations}\}      & s=2~ {\text{in}}~ (\IV)   
& \Delta n_{\text {T}}=2~ {\text in}~ (\III)     &  
\Delta n_{\text {T}}=2~ {\text in}~ (\V)   
\\ \hline \hline
1   & \text{A}    & {}         &            
 \\ \hline
1   & \text{B}    & {\text c}           &             
\\ \hline
0   & \text{C}    & {\text d}          & \alpha_{14}          
\\ \hline
1   & \text{D}   & {\text b}            &      
\\ \hline
1   & \text{E}    & {\text a}        &  \alpha_2,\alpha_7,
\alpha_{11},\alpha_{12},\alpha_{13},\alpha_{17}       
\\ \hline
2   &              &  {\text e}          &\alpha_1,\alpha_4,\alpha_8,
\alpha_9,\alpha_{15},\alpha_{16}     
\\ \hline
2   &              &  {\text f}          &\alpha_5,\alpha_6         
\\ \hline
2   &             &  {\text g}            &      
\\ \hline
1   &              &  {\text h}            &      
\\ \hline
3   &              &               & \alpha_{10}     
\\ \hline
1   &              &               & \alpha_{18}     
\\ \hline
0   &              &               &\alpha_{ 3}      
\\ \hline
\end{array}
\]
\caption{ Identification of phases. The phases 
in the same line are the same by criterion 1    }
\end{table}
}
\small{
\begin{table}[h]
\[
\begin{array}{|c|c|c|}\hline 
 {\rm phase} & c_2 \cdot  J_i  &{\rm equivalent ~phase}
\\ \hline 
A            & \{72,48,24,36,72'\}   & g,\alpha_{10} 
\\ \hline
a=E            & \{72,82,24,92,36\}  & h  \\ \hline
b=D            & \{72,92,24,82,48\}    & f,\alpha_{3}
\\ \hline
c=B            & \{72,82,24,36,48\} & e  
\\ \hline
d=C            & \{72,82,92,102,36\} &\alpha_{18}
\\ \hline
e            & \{72,36,24,24,82\}  & c=B     
\\ \hline
f            & \{72,82,24,24,92\}  &  b=D,\alpha_{3} 
\\ \hline
g            & \{72,24,36,24,48\}  & A,\alpha_{10}     
\\ \hline
h            & \{36,24,82,92,82'\}  &a=E        
\\ \hline
\alpha_{3}   & \{82,82,72,24,24\}  &   f,b=D   
\\ \hline
\alpha_{10}  & \{72,36,24,24,24\}   & A,g    
\\ \hline
\alpha_{18}  & \{92,92',82,72,36\}    & d=C    
\\ \hline
\end{array}
\]
\caption{
The relation of phases by criterion 2.
Phases in the same line denote the equivalent phases. 
There are only five phases, A,B,C,D, and E  of 
(IV) in {\bf HLY} model  which are  
topologically  non equivalent. }
\end{table}
}
\end{document}